\documentclass[runningheads]{llncs}

\usepackage{amsmath}
\usepackage{amsfonts}
\usepackage{algorithm}
\usepackage{algpseudocode}
\usepackage{dsfont}
\usepackage{algpseudocode}

\newcommand{\E}{\mathbb{E} }
\newcommand{\1}[1]{\mathds{1}_{#1}}
\newcommand*\diff{\mathop{}\!\mathrm{d}}

\newcommand{\Class}[1]{\mathcal{P}\left(#1\right) }

\newcommand{\Prc}[1]{\,\mathcal{P}_i\left[#1\right]}

\newenvironment{myproof}{{\it Proof.\ }}{\hfill$\Box$\vspace{3mm}}

\newcommand{\mysone}{n^{\frac{\alpha+4}{6(\alpha+1)}}}
\newcommand{\mystwo}{n^{\frac{11\alpha^2+3\alpha-2}{12\alpha(\alpha+1)}}}
\newcommand{\myk}{n^{\frac{7\alpha+10}{12(\alpha+1)}}}

\allowdisplaybreaks[1]

\title{Satisfiability Threshold for Power Law Random 2-SAT in Configuration Model\thanks{This work was supported by an NSERC Discovery grant}}
\titlerunning{Satisfiability Threshold for Power Law Random 2-SAT}
\author{Oleksii Omelchenko\inst{1} \and Andrei A. Bulatov\inst{1}}
\authorrunning{O.Omelchenko and A.Bulatov}
\institute{Simon Fraser University\\
\email{\{oomelche,abulatov\}@sfu.ca}}

\begin{document}

\maketitle

\begin{abstract}
The Random Satisfiability problem has been intensively studied for decades. For a number of reasons the focus of this study has mostly been on the model, in which instances are sampled uniformly at random from a set of formulas satisfying some clear conditions, such as fixed density or the probability of a clause to occur. However, some non-uniform distributions are also of considerable interest. In this paper we consider Random 2-SAT problems, in which instances are sampled from a wide range of non-uniform distributions. 

The model of random SAT we choose is the so-called configuration model, given by a distribution $\xi$ for the degree (or the number of occurrences) of each variable. Then to generate a formula the degree of each variable is sampled from $\xi$, generating several \emph{clones} of the variable. Then 2-clauses are created by choosing a random paritioning into 2-element
sets on the set of clones and assigning the polarity of literals at random. 

Here we consider the random 2-SAT problem in the configuration model for power-law-like distributions $\xi$. 	More precisely, we assume that $\xi$ is such that its right tail $F_{\xi}(x)$ satisfies the conditions $W\ell^{-\alpha}\le F_{\xi}(\ell)\le V\ell^{-\alpha}$ for some constants $V,W$. 	The main goal is to study the satisfiability threshold phenomenon depending on the parameters $\alpha,V,W$. We show that a satisfiability threshold exists and is determined by a simple relation between the first and second moments of $\xi$. 

\end{abstract}

\section{Introduction}
The Random Satisfiability problem (Random SAT) and its special cases Random $k$-SAT as a model of `typical case' instances of SAT has been intensively studied for decades. Apart from algorithmic questions related to the Random SAT, much attention has been paid to such problems as satisfiability thresholds and the structure of the solution space. The most widely studied model of the Random $k$-SAT is the uniform one parametrized by the (expected) density or clause-to-variable ratio $\varrho$ of input formulas. Friedgut in \cite{Friedgut99:sharp} proved that depending on the parameter $\varrho$ (and possibly the number of variables) Random $k$-SAT exhibits a sharp satisfiability threshold: a formula of density less than a certain value $\varrho_0$ (or possibly $\varrho_0(n)$) is satisfiable with high probability, and if the density is greater than $\varrho_0$, it is unsatisfiable with high probability. Moreover, a recent work of Friedrich and Rothenberger~\cite{Friedrich18Sharp}, which may be regarded as an extension of Fridgut's result to non-uniform random SAT instances, shows that if a distribution of variable's occurence in random formulas satisfies some criteria, then such formulas must undergo a sharp satisfiability threshold.

Hence, an impressive line of research aims at locating the satisfiability threshold for each random generating model. This includes more and more sophisticated methods of algorithms analysis \cite{Achlioptas01:lower,Coja-Oghlan10:better,%
	Coja-Oghlan13:going,Ding15:proof} and applications of the second moment method \cite{Achlioptas06:random} to find lower bounds, and a variety of probabilistic and proof complexity tools to obtain upper bounds \cite{Dubios97:general,Kirousis98:approximating,Boufkhad03:typical}. In the case of sufficiently large $k$ the exact location of the satisfiability threshold was identified by Ding, Sly and Sun \cite{Ding15:proof}. The satisfiability threshold and the structure of random $k$-CNFs received special attention for small values of $k$, see \cite{Chvatal92:mick,Vega01:random,Goerdt96:threshold} for $k=2$, and \cite{Diaz09:satisfiability,Kaporis06:probabilistic,Kirousis98:approximating} for $k=3$.

The satisfiability threshold phenomenon turned out to be closely connected with algorithmic properties of the Random SAT, as well as with the structure of its solution space. Experimental and theoretical results \cite{Selman96:generating,Cook96:finding}
demonstrate that finding a solution or proving unsatisfiability is hardest around the satisfiability threshold. The geometry of the solution space also exhibits phase transitions not far from the satisfiability threshold, related to various clustering properties \cite{Krzakala07:Gibbs}. This phenomenon has been exploited by applications of methods from statistical physics that resulted in some of the most efficient algorithms for Random SAT with densities around the satisfiability threshold \cite{Mezard02:analytic,Braunstein05:survey}.

Random $k$-SAT can be formulated using one of the three models whose statistical properties are very similar. In the model with fixed density $\varrho$, one fixes $n$ distinct propositional variables $v_1,\dots,v_n$ and then chooses $\varrho n$ $k$-clauses uniformly at random \cite{Franco83:probabilistic,Selman96:generating}. Alternatively, for selected variables every possible  $k$-clause is included with probability tuned up so that the expected number of clauses equals $\varrho n$. Finally, Kim \cite{Kim04:poisson} showed that one can also use the configuration model, which he called Poisson Cloning model. In this model for each variable $v_i$ we first select a positive integer $d_i$ accordingly to the Poisson distribution with expectation $k\varrho$, the degree of the variable. Then we create $d_i$ \emph{clones} of variable $v_i$, and choose $(d_1+\dots+d_n)/k$ random $k$-element subsets of the set of clones, then converting them into clauses randomly. The three models are largely equivalent and can be used whichever suits better to the task at hand.
  
The configuration model opens up a possibility for a wide range of different distributions of $k$-CNFs arising from different degree distributions. Starting with any random variable $\xi$ that takes positive integer values one obtains a distribution $\Phi(\xi)$ on $k$-CNFs as above using $\xi$ in place of the Poisson distribution. Note that $\xi$ may depend on $n$, the number of variables, and even be different for different variables. One `extreme' case of such a distribution is Poisson Cloning described above. Another case is studied by Cooper, Frieze, and Sorkin \cite{Cooper2007}. In their case each variable of a 2-SAT instance has a prescribed degree, which can be viewed as assigning a degree to every variable according to a random variable that only takes one value. We will be often returning to that paper, as our criterion for a satisfiability threshold is a generalization of that in \cite{Cooper2007}. Boufkhad et al.\ \cite{Boufkhad05:regular} considered another case of this kind --- regular Random $k$-SAT. 

In this paper we consider Random 2-SAT in the configuration model given by distribution $\Phi(\xi)$, where $\xi$ is distributed according to the power law  distribution in the following sense. Let $F_{\xi}(\ell)=\Pr[\xi \ge \ell]$ denote the tail function of a positive integer valued  random variable $\xi$. We say that $\xi$ is distributed according to the power law with parameter $\alpha$ if there exist constants $V,W$ such that 
\begin{equation}
W\ell^{-\alpha}\le F_{\xi}(\ell)\le V\ell^{-\alpha}.
\label{equ:tail-condition}
\end{equation}

Power law type distributions have received much attention. They have been widely observed in natural phenomena \cite{Newman05:power,Clauset09:power}, as well as in more artificial structures such as networks of various kinds \cite{Barabasi99:emergence}. Apart from the configuration model, graphs (and therefore 2-CNFs) whose degree sequences are distributed accordingly to a power law of some kind can also be generated in a number of ways. These include preferential attachment \cite{Aiello01:random,Barabasi99:emergence,Bollobas01:degree,%
Bollobas02:mathematical}, hyperbolic geometry \cite{Krioukov10:hyperbolic}, and others \cite{Ansotegui09:structure,Ansotegui09:towards}. Although the graphs resulting from all such processes satisfy the power law distributions of their degrees, other properties can be very different. We will encounter the same phenomenon in this paper. 

The approach most closely related to this paper was suggested by Ansotegui et al.\ \cite{Ansotegui09:structure,Ansotegui09:towards}. Given the number of variables $n$, the number of clauses $m$, and a parameter $\beta$, the first step in their construction is to create $m$ $k$-clauses without naming the variables. Then for every variable-place $X$ in every clause, $X$ is assigned to be one of the variables $v_1,\dots, v_n$ according to the distribution
\[
\Pr[X=v_i,\beta,n]=\frac{i^{-\beta}}{\sum_{j=1}^n j^{-\beta}}.
\]
Ansotegui et al.\ argue that this model often well matches the experimental results on industrial instances, see also \cite{AnsoteguiCarlos2016CSiI,Giraldez-CruJesus2016GSiw,Ansotegui08:measuring}. Interesting to note that although the model studied in these papers differ from the configuration model, it exhibits the same criterion of unsatisfiability $\E K^2 > 3\E K$, where $K$ is the r.v. that governs the number of times a variable appears in 2-SAT formula $\phi$~\cite{Levy}.  

The satisfiability threshold of this model has been studied by Friedrich et al.\ in \cite{Friedrich17:bounds}. Since the model has two parameters, $\beta$ and $r=m/n$, the resulting picture is  complicated. Friedrich et al.\ proved that a random CNF is unsatisfiable with high probability if $r$ is large enough (although constant), and if $\beta<\frac{2k-1}{k-1}$. If $\beta\ge\frac{2k-1}{k-1}$, the formula is satisfiable with high probability provided $r$ is smaller than a certain constant. The unsatisfiability results in \cite{Friedrich17:bounds} are mostly proved using the local structure of a formula.

In this paper we aim at a similar result for Random 2-SAT in the configuration model. Although the configuration model has only one parameter, the overall picture is somewhat more intricate, because there are more reasons for unsatisfiability than just the local structure of a formula. We show that for 2-SAT the parameter $\alpha$ from the tail condition (\ref{equ:tail-condition}) is what decides the satisfiability of such CNF. The main result of this paper is a satisfiability threshold given by the following

\begin{theorem}\label{the:main-intro}
Let $\phi$ be a random 2-CNF in the configuration model, such that the number of occurrences of each variable in $\phi$ is an independent copy of the random variable $\xi$, satisfying the tail condition (\ref{equ:tail-condition}) for some $\alpha$. 
Then for $n\to \infty$
	\[
	\Pr[\, \phi \text{ is satisfiable }] = 	\begin{cases}
	0, \text{ when } &  0<\alpha<2\\
	0, \text{ when } & \alpha=2 \text{ or } \E \xi^2>3\E \xi,\\
	1, \text{ when}  & \E \xi^2<3\E \xi.  
	\end{cases}
	\]
\end{theorem}

In the first case of Theorem~\ref{the:main-intro} we show that 
$\phi$ is unsatisfiable with high probability due to very local
structure of the formula, such as the existence of variables
of sufficiently high degree. Moreover, same structures persist with
high probability in $k$-CNF formulas for \textit{any} $k \geq 2$ obtained 
from the configuration model, when $\alpha<\frac{k}{k-1}$. 

In the remaining cases we apply the approach of Cooper, Frieze, and Sorkin \cite{Cooper2007}. 
It makes use of the structural characterization of unsatisfiable
2-CNFs: a 2-CNF is unsatisfiable if and only if it contains so-called
\emph{contradictory paths}. If $\E \xi^2<3\E \xi$ we prove 
that w.h.p.\ formula $\phi$ does not have long paths, and 
contradictory paths are unlikely to form. If $\E \xi^2>3\E \xi$,
we use the analysis of the dynamics of the growth of $\phi$ 
to show that contradictory paths appear w.h.p.
However, the original method by Cooper et al.\ only works with 
strong restrictions on the maximal degree of variables that 
are not affordable in our case, and so it requires substantial 
modifications.

\section{Notation and preliminaries}
\label{sec:prelims}

We use the standard terminology and notation of variables, positive and negative literals, clauses and 2-CNFs, and degrees of variables. The degree of variable $v$ will be denoted by $deg(v)$, or when our CNF contains only variables $v_1,\dots,v_n$, we use $d_i=deg(v_i)$. By $C(\phi)$ we denote the set of clauses in $\phi$, while  by $V(\phi)$ and $L(\phi)$ we denote the sets of variables and literals of $\phi$ respectively.
Let $d_{i}^{+}$ denote the number of occurrences of $v_i$ as a positive literal (or the number of the literal $v_i$), and let $d_{i}^{-}$ denote the number of occurrences of the literal $\bar{v}_i$. 

\subsection{Configuration model}

We describe the configuration model for $k$-CNFs, but will only use it for $k=2$, see also \cite{Kim04:poisson}. In the configuration model of $k$-CNFs with $n$ variables $v_1,\dots, v_n$ we are given a positive integer-valued random variable (r.v.) $\xi$ from which we sample independently $n$ integers $\{ d_i \}_{i=1}^{n}$. Then $d_i$ is the degree of $v_i$, that is, the number of occurrences of $v_i$ in the resulting formula $\phi$. Each occurrence of $v_i$ in $\phi$ we call a \emph{clone} of $v_i$. Hence, $d_i$ is the number of clones of $v_i$. Then we sample $k$-element sets of clones from the set of all clones without replacement. Finally, every such subset is converted into a clause by choosing the polarity of every clone in it uniformly at random. If the total number of clones is not a multiple of $k$, we discard the set and repeat the procedure. Algorithm~\ref{configuration-model} gives a more precise description of the process. We will sometimes say that a clone $p$ is associated with variable $v$ if $p$ is a clone of $v$. In a similar sense we will say a clone associated with a literal if we need to emphasize the polarity of the clone. 

\begin{algorithm}
\caption{Configuration Model $\mathbb{C}_{n}^{k}(\xi)$}
\label{configuration-model}
\begin{algorithmic}[1]
\Procedure{SampleCNF}{$n,k,\xi$}
    \State Form a sequence of $n$ numbers $\{d_{i}\}_{i=1}^{n}$ each sampled independently from $\xi$
    \If{$S_{n} := \sum_{i=1}^{n}d_{i}$ is not a multiple of $k$} 
    	\State {discard the sequence, and go to step 2}
    \EndIf
    \State Otherwise, introduce multi-set $S\leftarrow\bigcup\limits_{i=1}^{n}\underbrace{\{ v_i, v_i, \dotsc, v_i \}}_{d_i \text{ times}}$
    \State Let $\phi \leftarrow \emptyset$
    \While {$S \neq \emptyset$}
    	\State Pick u.a.r. $k$ elements $\{v_{1}, v_{2}, \dots, v_{k}\}$ from $S$ without replacement
    	\State Let $C \leftarrow \{v_{1}, v_{2}, \dots, v_{k}\}$
    	\State $S \leftarrow S - C$
    	\State Negate each element in $C$ u.a.r with probability 1/2 
    	\State $\phi \leftarrow \phi \cup C$
    \EndWhile
    \State\Return $\phi$
\EndProcedure
\end{algorithmic}
\end{algorithm}

We will denote a random formula $\phi$ obtained from $\mathbb{C}_{n}^{k}(R)$ by $\phi \sim \mathbb{C}_{n}^{k}(R)$.
Clearly, formulas $\phi \sim \mathbb{C}_{n}^{k}(R)$ are defined over a set of $n$ Boolean variables, 

\subsection{Power law distributions}

We focus our attention on the configuration model $\mathbb{C}_{n}^{k}(\xi)$, in which every variable is an i.i.d. copy of the random variable $\xi$ having power-law distribution. In this paper we define such distributions through the properties of their tail functions. If $\xi$ is an integer-valued r.v., its tail function is defined to be $F_\xi(\ell)=\Pr[\xi\geq \ell]$, where $\ell \geq 1$.

\begin{definition}
\label{definition}
An integer-valued positive r.v. $\xi$ has power-law probability distribution, if 
$
F_\xi(\ell) =  \Theta\left( \ell^{-\alpha} \right),
$
where $\alpha > 0$. We denote this fact as $\xi\sim \Class{\alpha}$.
\end{definition}  

Clearly, if $\xi\sim \Class{\alpha}$, then there exist constants $V,W>0$, such that
$
W\,\ell^{-\alpha} \leq F_\xi(\ell) \leq V\, \ell^{-\alpha},
$
for every $\ell \geq 1$.

The existence of the moments of $\xi\sim \Class{\alpha}$ depends only on $\alpha$.

\begin{lemma}
\label{lemma:moments}
Let $\xi\sim \Class{\alpha}$. Then $\E\xi^{m} < \infty$ iff $0 < m < \alpha$.
\end{lemma}

\begin{myproof}
From the definition of $m$-th moment of a strictly positive r.v. $X$, we have
\begin{align*}
\mathbb{E} X^m  &= \sum_{\ell \geq 1}k^{m}\Pr\left[X = \ell\right] \nonumber \\
                &= \sum_{\ell \geq 1}k^{m}\Big(\Pr\left[X \geq \ell\right]-\Pr\left[X \geq \ell-1\right]\Big) \nonumber \\
                &= \sum_{\ell \geq 1}\ell^{m}\Pr\left[X \geq \ell\right]-\sum_{\ell \geq 2}(\ell-1)^m\Pr\left[X \geq \ell\right] \nonumber \\
                &= \Pr\left[X \geq 1\right] + \sum_{\ell \geq 2}\ell^{m}\Pr\left[X \geq \ell\right]-\sum_{\ell \geq 2}(\ell-1)^m\Pr\left[X \geq \ell\right]\\
			    &= 1 + \sum_{\ell \geq 2}\Big(\ell^{m}-(\ell-1)^{m}\Big)\Pr\left[X \geq \ell\right]\\
			    &= \sum_{\ell \geq 1}\Big(\ell^{m}-(\ell-1)^{m}\Big)\Pr\left[X \geq \ell\right]\\
                &= \sum_{\ell \geq 1}\Big(\ell^{m}-\sum_{i=0}^{m}\binom{m}{i}\ell^{m-i}(-1)^{i}\Big)\Pr\left[X \geq \ell\right]\\
                &= \sum_{\ell \geq 1}\sum_{i=1}^{m}\binom{m}{i}\ell^{m-i}(-1)^{i+1}\Pr\left[X \geq \ell\right]\\
                &= \sum_{i=1}^{m}\binom{m}{i}(-1)^{i+1}\sum_{\ell \geq 1}\ell^{m-i}\Pr\left[X \geq \ell\right].
\end{align*}
However, since $X \sim  \Class{\alpha}$, then for any $\ell \geq 1$
$$
\Pr[X \geq \ell] = F_{X}(\ell) \leq V\,\ell^{-\alpha}.
$$
Therefore,
\begin{align*}
\mathbb{E} X^m  &= \sum_{i=1}^{m}\binom{m}{i}(-1)^{i+1}\sum_{\ell \geq 1}\ell^{m-i}\Pr\left[X \geq \ell\right]\\
  &\leq \sum_{i=1}^{m}\binom{m}{i}(-1)^{i+1}\sum_{\ell \geq 1}\ell^{m-i}V\,\ell^{-\alpha}\\
  & = V\sum_{i=1}^{m}\binom{m}{i}(-1)^{i+1}\sum_{\ell \geq 1}\ell^{m-i-\alpha},
\end{align*}
which is finite iff $0<m<\alpha$, and the result follows.
\end{myproof}

We will write $\E\xi^{m}=\infty$ when the \textit{m}-th moment of some r.v. $\xi$ is not finite or does not exist. We will have to deal with cases when the second or even first moment of $\xi$ does not exist.

Nevertheless, we can obtain good bounds on useful quantities formed from such variables with a good level of confidence, despite the absence of expectation or variance. One such quantity is the sum of \textit{independent} variables drawn from $\Class{\alpha}$:
$
S_n=\sum_{i=1}^{n}\xi_i,
$     
where $\xi_i \sim \Class{\alpha}$. Note that $\xi_i$'s are not required to be identically distributed. They can come from different distributions, as long as their right tail can be bounded with some power-law functions with exponent $\alpha$. But we do require their independence.

The next two theorems provide bounds on the values of $S_n$, depending on $\alpha$ in a slightly more general case of r.vs.\ admitting negative values. 

\begin{theorem}[Corollary 1 from~\cite{Omelchenko18:concentration}]
\label{theorem:sum-1}
Let $S_n=\sum_{i=1}^{n}\xi_i$, where $\xi_i$'s are independent integer-valued random variables, with
$$
\Pr\left[ \xi_i \geq \ell  \right] \leq V\,\ell^{-\alpha}, \quad \text{ and } \quad \Pr\left[ \xi_i \leq -\ell  \right] \leq V\,\ell^{-\alpha},
$$
where $V>0$ and $0<\alpha \leq 1$ are constants. Then w.h.p. $S_n \leq C \, n^{\frac{1}{\alpha}}$, where  $C>0$ is some constant.
\end{theorem}  

As for the second theorem, we deal with a similar sum of random variables, but each variable's tail can be majorized with a power-law function with exponent $\alpha>1$. Then, as it follows from Lemma~\ref{lemma:moments}, such variables have finite expectation, and due to the linearity of expectation, the sum itself has well defined mean value. 

\begin{theorem}[Corollary 5 from~\cite{Omelchenko18:concentration}]
\label{theorem:sum-2}
Let $S_n=\sum_{i=1}^{n}\xi_i$, where $\xi_i$'s are independent integer-valued random variables, with
$$
\Pr\left[ \xi_i \geq \ell  \right] \leq V\,\ell^{-\alpha}, \quad \text{ and } \quad \Pr\left[ \xi_i \leq -\ell  \right] \leq V\,\ell^{-\alpha},
$$
where $V>0$ and $\alpha > 1$ are constants. Then w.h.p. $S_n = \sum_{i=1}^{n}\E \xi_i + o(n)$.
\end{theorem}  

Hence, as the theorem states, when  $\xi_i$'s are independent r.vs. with power-law boundable tails with tail exponent $\alpha>1$, then the sum of such variables does not deviate much from its expected value.

Note that from now on we will deal with sctrictly positive power-law r.vs.\ $\xi_i$'s, hence, their expectation (given that it exists) is a positive constant. Then when $\alpha>1$, we have $S_n = \sum_{i=1}^{n}\E \xi_i + o(n) = (1+o(1))\,\sum_{i=1}^{n}\E \xi_i$.

Another important quantity we need is the maximum, $\Delta$, of the sequence of $n$ independent random variables (or the maximum degree of a CNF in our case). 

\begin{lemma}
\label{lemma:delta}	
Let $\Delta = \max\left( \xi_1,\, \xi_2, \, \cdots, \, \xi_n \right)$, where $\xi_i$'s are independent copies of an r.v. $\xi \sim \Class{\alpha}$ with $\alpha > 0$. Then w.h.p.
$
\Delta \leq C\,n^{1/\alpha},
$
where $C>0$ is some constant.
\end{lemma}

\begin{myproof}
Simple calculation shows that
\begin{align*}
\Pr\left[\Delta \geq x\right] &= \Pr\left[ \max\left(\xi_1, \xi_2, \cdots, \xi_n\right) \geq x \right]\\
& = \Pr\left[\bigcup_{i=1}^{n}\{ \xi_i \geq x \}\right]\\
& \leq \sum_{i=1}^{n} \Pr\left[\xi_i \geq x \right], \text{ by Union bound}\\
& =  \sum_{i=1}^{n} \Pr\left[\xi \geq x \right], \text{ since } \xi_i \overset{d}{=}\xi \\
& = nF_{\xi}(x).
\end{align*}

Since $\xi \sim \Class{\alpha}$, we have that $F_{\xi}(\ell) \leq V \, \ell^{-\alpha}$. Hence,
$$
\Pr\left[\Delta \geq x\right] \leq nF_{\xi}(x) \leq nVx^{-\alpha}.
$$
Then if $\Delta\geq n^{1/\alpha+\epsilon}$ for any $\epsilon>0$, we obtain
$$
\Pr\left[\Delta \geq n^{1/\alpha+\epsilon}\right] \leq  V\,n^{-\alpha\epsilon}=o(1),
$$
hence, we do not expect to see variables with such large degrees. Thus, we conclude that $\Delta \leq C\,n^{1/\alpha}$ holds w.h.p., and the lemma follows.
\end{myproof}

We will also need some bounds on the number of pairs of complementary clones of a variable $v_i$, that is, the value $d_i^+d_i^-$. By the definition of the configuration model 
$$
d_{i}^{+} \sim Bin\Big(\deg(v_{i}), 1/2\Big) \quad \text{ and } \quad d_{i}^{-}=\deg(v_i) - d_{i}^{+},
$$ 
where $Bin(n,p)$ is the Binomial distribution with $n$ trials and success probability $p$.


\begin{lemma}
\label{lemma:literal-tail}
Let $\xi$ be some positive integer-valued r.v., and let $d^{+} \sim Bin\Big(\xi, 1/2\Big)$, while $d^{-}=\xi - d^{+}$. Then
\[
F_{d^{+}}(\ell)=F_{d^{-}}(\ell) \leq F_{\xi}(\ell),
\quad {and} \quad 
F_{d^{+}d^{-}}(\ell) \leq 2F_{\xi}\left(\ell^{1/2}\right).
\]
\end{lemma}

\begin{myproof}
	Observe that the event $\{ d^{+} \geq \ell \}$ implies $\{ \xi \geq \ell \}$, therefore, $\{ d^{+} \geq \ell \} \subseteq \{ \xi \geq \ell \}$, and as a result
	$$
	F_{d^{+}}(\ell) = \Pr\left[ d^{+} \geq \ell \right] \leq \Pr\left[ \xi \geq \ell \right] = F_{\xi}(\ell),
	$$ 
	and since $d^{+} \overset{d}{=} d^{-}$, the first result follows. As for the number of pairs of complementary clones, we have 
	
	\begin{align}
	F_{d^{+}d^{-}}(\ell) &= \Pr\left[d^{+}d^{-} \geq \ell  \right] \nonumber \\
	&= \Pr\left[d^{+}d^{-} \geq \ell \,  | \, d^{+} \geq \ell^{1/2}\right]\Pr\left[d^{+} \geq \ell^{1/2}\right] \nonumber \\
	& \qquad  + \Pr\left[d^{+}d^{-} \geq \ell \, | \, d^{+} < \ell^{1/2}\right]\Pr\left[d^{+} < \ell^{1/2}\right] \nonumber \\
	& \leq \Pr\left[d^{+} \geq \ell^{1/2}\right] + \Pr\left[d^{+}d^{-} \geq \ell \, | \, d^{+} < \ell^{1/2}\right] \nonumber \\
	& \leq \Pr\left[d^{+} \geq \ell^{1/2}\right] + \Pr\left[d^{-} \geq \ell^{1/2} \, | \, d^{+} < \ell^{1/2}\right] \label{eq:last-line}.
	\end{align}  
	
	We have already established that $\Pr\left[d^{+} \geq \ell^{1/2}\right]  \leq F_{\xi}\left(\ell^{1/2}\right)$. As for the second probability in~\eqref{eq:last-line}, it is bounded in a similar manner. Since the event $\{ d^{-} \geq \ell^{1/2} \, | \, d^{+} < \ell^{1/2} \}$ implies $\{ \xi \geq \ell^{1/2} \}$, it follows
	$$
	\Pr\left[d^{-} \geq \ell^{1/2} \, | \, d^{+} < \ell^{1/2}\right] \leq \Pr\left[\xi \geq \ell^{1/2}\right] = F_{\xi}\left(\ell^{1/2}\right).
	$$ 
	Hence, after combining the two probabilities together, we obtain that
	$$
	F_{d^{+}d^{-}}(\ell) \leq \Pr\left[d^{+} \geq \ell^{1/2}\right] + \Pr\left[d^{-} \geq \ell^{1/2} \, | \, d^{+} < \ell^{1/2}\right] \leq 2F_{\xi}\left(\ell^{1/2}\right),
	$$
	and the lemma follows.
\end{myproof}

Hence, for $\xi \sim \Class{\alpha}$, we have 
\begin{corollary}
\label{cor:literal-tail}
Let $\xi \sim \Class{\alpha}$, where $\alpha>0$, be some positive integer-valued r.v., and let $d^{+} \sim Bin\Big(\xi, 1/2\Big)$, while $d^{-}=\xi - d^{+}$. Then
\begin{align}
F_{d^{+}}(\ell) =F_{d^{-}}(\ell) & \leq V \, \ell^{-\alpha},\label{eq:literal-tail}\\
F_{d^{+}d^{-}}(\ell) & \leq 2V \, \ell^{-\alpha/2}. \label{eq:pair-tail}
\end{align}
\end{corollary}

The expectations of $d^+$ and $d^-$ are easy to find: $\E d^{+} = \E d^{-} = \frac{\E\xi}{2}$. However, the expected value of $d^{+}d^{-}$ requires a little more effort.

\begin{lemma}
\label{lemma:dd}
Let $\xi$ be some positive integer-valued r.v., and let $d^{+} \sim Bin\Big(\xi, 1/2\Big)$, while $d^{-}=\xi - d^{+}$. Then 
$
\E \left[ d^{+}d^{-} \right] = \frac{\E\xi^2 - \E\xi}{4}.
$
\end{lemma}

\begin{myproof}
From the definition of the expected value, and the way quantities $d^{+}$ and $d^{-}$ are calculated, it follows that	
\begin{align*}
\E\left[d^{+} d^{-}\right] &=\sum_{\ell=1}^{\infty}\sum_{d=0}^{\ell}d\left(\ell-d\right)\Pr\left[d^{+}=d \,|\, \xi=\ell \right]\,\Pr\left[\xi=\ell\right]\\
&= \sum_{\ell=1}^{\infty}\sum_{d=0}^{\ell}d\left(\ell-d\right)\Pr\left[Bin(\ell,1/2)=d \,|\, \xi=\ell \right]\,\Pr\left[\xi=\ell\right]\\
&= \sum_{\ell=1}^{\infty}\sum_{d=0}^{\ell}d\left(\ell-d\right)\binom{\ell}{d}\frac{1}{2^\ell}\,\Pr\left[\xi=\ell\right]\\
&= \sum_{\ell=1}^{\infty}\sum_{d=0}^{\ell}\ell\,d\binom{\ell}{d}\frac{1}{2^\ell}\,\Pr\left[\xi=\ell\right]-\sum_{\ell=1}^{\infty}\sum_{d=0}^{\ell}d^{2}\binom{\ell}{d}\frac{1}{2^\ell}\,\Pr\left[\xi=\ell\right]\\
&= \sum_{\ell=1}^{\infty}\frac{\ell\Pr\left[\xi=\ell\right]}{2^\ell}\sum_{d=0}^{\ell}d\binom{\ell}{d}-\sum_{\ell=1}^{\infty}\frac{\Pr\left[\xi=\ell\right]}{2^\ell}\sum_{d=0}^{\ell}d^{2}\binom{\ell}{d}.
\end{align*}

Next, we apply two well-known relations
$$
\sum_{j=0}^{n}j\binom{n}{j} = n2^{n-1} \quad \text{ and } \quad \sum_{j=0}^{n}j^2\binom{n}{j} = (n+n^2)2^{n-2},
$$
to get 
\begin{align*}
\E\left[d^{+} d^{-}\right] &= \sum_{\ell=1}^{\infty}\frac{\ell\Pr\left[\xi=\ell\right]}{2^\ell}\sum_{d=0}^{\ell}d\binom{\ell}{d}-\sum_{\ell=1}^{\infty}\frac{\Pr\left[\xi=\ell\right]}{2^\ell}\sum_{d=0}^{\ell}d^{2}\binom{\ell}{d}\\
&= \sum_{\ell=1}^{\infty}\frac{\ell\Pr\left[\xi=\ell\right]}{2^\ell} \cdot \ell 2^{\ell-1}-\sum_{\ell=1}^{\infty}\frac{\Pr\left[\xi=\ell\right]}{2^\ell}\cdot (\ell+\ell^2)2^{\ell-2}\\
&= \frac{1}{2}\sum_{\ell=1}^{\infty}\ell^2\Pr\left[\xi=\ell\right]-\frac{1}{4}\sum_{\ell=1}^{\infty}\ell\Pr\left[\xi=\ell\right]-\frac{1}{4}\sum_{\ell=1}^{\infty}\ell^2\Pr\left[\deg(v_i)=d\right]\\
&= \frac{\E\xi^2-\E\xi}{4},
\end{align*}
and the proof is finished.
\end{myproof}

We use $T_n = \sum_{i=1}^{n}d_{i}^{+} d_{i}^{-}$ to denote the total number of pairs of complementary clones, i.e. the sum of unordered pairs of complementary clones over all $n$ variables,.

Note, that when $\alpha > 2$, the r.v.\ $d_{i}^{+} d_{i}^{-}$ has finite expectation due to Lemma~\ref{lemma:moments}. Then by Theorem~\ref{theorem:sum-2} w.h.p. holds
$$
T_n = (1+o(1))\sum_{i=1}^{n}\E\left[d_{i}^{+} d_{i}^{-}\right].
$$

We finish this subsection with \textit{Azuma-like inequality} first appeared in~\cite{Cooper2007}, which will be used in the proofs. 
Informally, the inequality states that a discrete-time random walk $X=\sum_{i=1}^{n}X_{i}$ with positive drift, consisting of not necessary independent steps, each having a right tail, which can be bounded by a power function with exponent at least 1, is very unlikely to drop much below the expected level, given $n$ is large enough. Although, the original proof was relying on the rather artificial step of introducing a sequence of uniformly distributed random numbers, we figured out that the same result can be obtained by exploiting the tower property of expectation. 

\begin{lemma}[Azuma-like inequality]
\label{lemma:azuma}
Let $X=X_0 + \sum_{i=1}^{t}X_{i}$ be some random walk, such that $X_0 \geq 0$  is constant initial value of the process, $X_{i} \geq -a$, where $a>0$ is constant,  are bounded from below random variables, not necessary independent, and such that $\mathbb{E}[X_{i} \, | \, X_{1},\dots,X_{i-1}] \geq \mu > 0$ ($\mu$ is constant) and \newline $\Pr[X_{i} \geq \ell \, | \, X_{1},\dots,X_{i-1}] \leq V\,\ell^{-\alpha}$ for every $\ell \geq 1$ and constants $V>0$, $\alpha>1$. Then for any $0 < \varepsilon < \frac{1}{2}$, the following inequality holds
$$
\Pr\left[X \leq \varepsilon\mu t\right] \leq \exp\left( - \frac{t+X_0}{4\log^2 t}\mu^2\left(\frac{1}{2}-\varepsilon\right)^2 \right).
$$
\end{lemma}

\begin{myproof}
First, let us introduce ``truncated'' at $\delta:=\lfloor \log n \rfloor$ versions of the variables $X_i$, i.e.
$$
Y_i = X_i\cdot\1{X_i \leq \delta}.
$$
Then the conditional expectation for the new variables is
\begin{align*}
\E \left[Y_i \, | \, X_1, \dots, X_{i-1} \right] & = \E\left[ X_i\cdot\1{X_i \leq \delta}  \, | \, X_1, \dots, X_{i-1} \right]\\
& = \E\left[ X_i - X_i\cdot\1{X_i > \delta}  \, | \, X_1, \dots, X_{i-1}  \right]\\
& = \E \left[X_i  \, | \, X_1, \dots, X_{i-1} \right] - \E\left[ X_i\cdot\1{X_i > \delta}  \, | \, X_1, \dots, X_{i-1} \right]\\
& \geq \E \left[X_i \, | \, X_1, \dots, X_{i-1} \right] - \E\left[ X_i\cdot\1{X_i \geq \delta}  \, | \, X_1, \dots, X_{i-1} \right]\\
& \geq \mu - \E\left[ X_i\cdot\1{X_i \geq \delta}  \, | \, X_1, \dots, X_{i-1} \right],
\end{align*}
since $\E\left[X_i \, | \, X_1,\dots,X_{i-1}\right] \geq \mu$. Let us denote by $\Prc{A}$, where $A$ is some event,  the following function
$$
\mathcal{P}_i\left[A\right]  := \Pr\left[A \, | \, X_1,\dots, X_{i-1}\right].
$$
Then
\begin{align*}
\E \left[Y_i \, | \, X_1, \dots, X_{i-1} \right] & \geq \frac{\mu}{2} - \E\left[ X_i\cdot\1{X_i \geq \delta}  \, | \, X_1, \dots, X_{i-1} \right]\\
& = \mu - \sum_{\ell=\delta}^{\infty}\ell\Pr\left[X_i=\ell \, | \, X_1,\dots, X_{i-1}\right] \\
& = \mu - \sum_{\ell=\delta}^{\infty}\ell\Prc{X_i=k} \\
& = \mu - \sum_{\ell=\delta}^{\infty}\ell\Big(\Prc{ X_i \geq \ell}-\Prc{X_i \geq \ell+1}  \Big)\\
& = \mu - \Big(\sum_{\ell=\delta}^{\infty}\ell\Prc{X_i \geq \ell}-\sum_{\ell=\delta+1}^{\infty}(\ell-1)\Prc{X_i \geq \ell}\Big)\\
& = \mu -\delta\Prc{X_i \geq \delta}- \Big(\sum_{\ell=\delta+1}^{\infty}\ell\Prc{X_i \geq \ell}-\sum_{\ell=\delta+1}^{\infty}(\ell-1)\Prc{X_i \geq \ell}\Big)\\
& = \mu - \delta\Prc{X_i \geq \delta} -\sum_{\ell=\delta+1}^{\infty}\Pr\left[ X_i \geq \ell \, | \, X_1,\dots, X_{i-1} \right].
\end{align*}

Now recall that $\Prc{X_i \geq \delta}  = \Pr\left[X_i \geq \delta \, | \, X_1,\dots,X_{i-1}\right] \leq V\,\delta^{-\alpha}$. Hence,
\begin{align*}
\E \left[Y_i \, | \, X_1, \dots, X_{i-1} \right]  & \geq \mu -\delta\Prc{X_i \geq \delta} -\sum_{\ell=\delta+1}^{\infty}\Pr\left[ X_i \geq \ell \, | \, X_1,\dots, X_{i-1} \right]\\
& \geq \mu -V\,\delta^{1-\alpha}-V\sum_{\ell=\delta+1}^{\infty}\ell^{-\alpha}\\
& \geq \mu -V\,\delta^{1-\alpha}-V\int\limits_{\delta}^{\infty}x^{-\alpha} \diff x\\
& = \mu -V\,\delta^{1-\alpha}-V\frac{\delta^{1-\alpha}}{\alpha-1}\\
& = \mu - C\,\delta^{1-\alpha},
\end{align*}
where $C>0$ is constant.

Since $\delta = \lfloor \log n \rfloor$ and $\alpha>1$, we obtain
\begin{align*}
\E \left[Y_i \, | \, X_1, \dots, X_{i-1} \right]  & \geq \mu -C\,\delta^{1-\alpha} =  \mu - C\,\left(\lfloor\log n \rfloor\right)^{1-\alpha}  \geq \mu - o(1) \geq \frac{\mu}{2}.
\end{align*}

Next, notice that the above conditional expectation's  lower bound is irrelevant to the realizations of the r.vs. $X_1,\dots,X_{i-1}$. Thus, for \textit{any} trajectory of the random walk $\{x_j\}_{j=1}^{i-1}$, where $x_j \in D(X_j)$ is the element from the domain of the r.v. $X_j$, we have that $\E\left[Y_i \, | \, x_1,\dots,x_{i-1}\right] \geq \frac{\mu}{2}$. Hence, since $D(Y_i) \subseteq D(X_i)$, it follows that
\begin{align}
\E\left[Y_i \, | \, Y_1,\dots, Y_{i-1}\right] \geq \frac{\mu}{2}. \label{eq:azuma-mean}
\end{align}
As for the second moment, since $Y_i \leq \delta$, we have an obvious upper bound 
\begin{align}
\E \left[Y_i^2 \, | \, Y_1,\dots,Y_{i-1}\right] \leq \delta^2 \leq \log^2 n.\label{eq:azuma-var}
\end{align}	

Now, using the truncated variables, we introduce an auxiliary random process
$$
Y = X_0 + \sum_{i=1}^{t}Y_i.
$$

Then the probability that the original random walk $X$ will drop below the $\epsilon t\mu$ level, where $0<\epsilon<1/2$, is at most
\begin{align*}
\Pr\left[ X \leq \epsilon t\mu\right] &\leq \Pr\left[Y \leq \epsilon t\mu\right], \text{ since } Y \leq X\\
& = \Pr\left[ e^{-\lambda Y} \geq e^{-\lambda \epsilon t\mu} \right], \text{ for any } \mu > 0\\
& \leq e^{\lambda \epsilon t\mu} \E e^{-\lambda Y}, \text{ by Markov's inequality}\\
& = e^{\lambda \epsilon t\mu-\lambda X_0}\E e^{-\lambda\sum_{i=1}^{t}Y_i}\\
& = e^{\lambda \epsilon t\mu-\lambda X_0}\E \left[\prod_{i=1}^{t}e^{-\lambda Y_i}\right].
\end{align*}

Next, we apply the tower property of expectation, that is for any two random variables $A$ and $B$ defined over the same probability space and $\E |A| < \infty$, the following holds (subscript indicates over which variable calculation of expectation is performed)
$$
\E A =\E_A \left[A\right] = \E_B\Big[\E_A \left[A\,|\,B\right]\Big].
$$ 
Thus, we have
\begin{align*}
\Pr\left[ X \leq \epsilon t\mu\right] & \leq e^{\lambda \epsilon t\mu-\lambda X_0}\E \left[\prod_{i=1}^{t}e^{-\lambda Y_i}\right]\\
& = e^{\lambda \epsilon t\mu-\lambda X_0}\E_{Y_1,\dots,Y_t} \left[\prod_{i=1}^{t}e^{-\lambda Y_i}\right]\\
& = e^{\lambda \epsilon t\mu-\lambda X_0}\E_{Y_1,\dots,Y_{t-1}} \left[\E_{Y_t}\left[\prod_{i=1}^{t}e^{-\lambda Y_i} \, | \, Y_1,\dots,Y_{i-1}\right]\right].
\end{align*}

Consider the innermost expectation. Since we condition it over variables $Y_1,\dots,Y_{t-1}$, we consider such variables as given (or constant). Hence,
\begin{align}
\Pr\left[ X \leq \epsilon t\mu\right] & \leq e^{\lambda \epsilon t\mu-\lambda X_0}\E_{Y_1,\dots,Y_{t-1}} \left[\E_{Y_t}\left[\prod_{i=1}^{t}e^{-\lambda Y_i} \, | \, Y_1,\dots,Y_{i-1}\right]\right] \nonumber \\
& = e^{\lambda \epsilon t\mu-\lambda X_0}\E_{Y_1,\dots,Y_{t-1}} \left[\prod_{i=1}^{t-1}e^{-\lambda Y_i}\cdot\E_{Y_t}\left[e^{-\lambda Y_t} \, | \, Y_1,\dots,Y_{i-1}\right]\right]. \label{eq:tmp-3}
\end{align}

Now, since $Y_i \geq -a$ just like the original variables $X_i$'s, we can upper bound the inner expectation by applying the well-known inequality
$$
e^{-x} \leq 1-x+x^2,
$$
which is valid for every $x \geq -1$. Hence, by restricting $0<\lambda \leq 1/a$, we have
\begin{align*}
\E_{Y_t}\left[e^{-\mu Y_t} \, | \, Y_1,\dots,Y_{t-1}\right] &\leq \E_{Y_t}\left[1 -\lambda Y_t + \lambda^2 Y_t^2 \, | \, Y_1,\dots,Y_{t-1}\right]\\
& = 1 - \lambda\E_{Y_t}\left[ Y_t \, | \, Y_1,\dots,Y_{t-1}\right]+ \lambda^2 \E\left[Y_t^2 \, | \, Y_1,\dots,Y_{t-1}\right]\\
& \leq 1 - \lambda\frac{\mu}{2} +\lambda^2\log^2 t,
\end{align*}
since $\E\left[ Y_i \, | \, Y_1,\dots,Y_{i-1}\right] \geq \frac{\mu}{2}$~\eqref{eq:azuma-mean} and $\E\left[ Y_i^2 \, | \, Y_1,\dots,Y_{i-1}\right] \leq \log^2 t$~\eqref{eq:azuma-var}. Therefore, we obtain that
$$
\E_{Y_t}\left[e^{-\mu Y_t} \, | \, Y_1,\dots,Y_{t-1}\right] \leq 1 - \lambda\frac{\mu}{2} +\lambda^2\log^2 t \leq \exp\left( - \lambda\frac{\mu}{2} +\lambda^2\log^2 t \right).
$$

Thus, the probability in~\eqref{eq:tmp-3} is upper bounded as
\begin{align*}
\Pr\left[ X \leq \epsilon t\mu\right] & \leq e^{\lambda \epsilon t\mu-\lambda X_0}\E_{Y_1,\dots,Y_{t-1}} \left[\prod_{i=1}^{t-1}e^{-\lambda Y_i}\cdot\E_{Y_t}\left[e^{-\lambda Y_t} \, | \, Y_1,\dots,Y_{t-1}\right]\right]\\
& \leq e^{\lambda \epsilon t\mu-\lambda X_0}\E_{Y_1,\dots,Y_{t-1}} \left[\prod_{i=1}^{t-1}e^{-\lambda Y_i}\cdot \exp\left( - \lambda\frac{\mu}{2} +\lambda^2\log^2 t \right)\right]\\
& \leq e^{\lambda \epsilon t\mu-\lambda X_0}\cdot \exp\left( - \lambda\frac{\mu}{2} +\lambda^2\log^2 t \right)\E_{Y_1,\dots,Y_{t-1}} \left[\prod_{i=1}^{t-1}e^{-\lambda Y_i}\right].
\end{align*}
Repeating the same process inductively for another $t-1$ times, we obtain
\begin{align*}
\Pr\left[ X \leq \epsilon t\mu\right] & \leq e^{\lambda \epsilon t\mu-\lambda X_0}\cdot \exp\left( - \lambda\frac{\mu}{2} +\lambda^2\log^2 t \right)\E_{Y_1,\dots,Y_{t-1}} \left[\prod_{i=1}^{t-1}e^{-\lambda Y_i}\right]\\
& \leq e^{\lambda \epsilon t\mu-\lambda X_0}\cdot \exp\left( - \lambda t \frac{\mu}{2} +\lambda^2 t \log^2 t \right)\\
& = \exp\left( \lambda \epsilon t\mu -\lambda X_0 - \lambda t \frac{\mu}{2} +\lambda^2 t \log^2 t \right).
\end{align*}

Fix $\lambda = \frac{\mu}{2\log^2 t}\left(\frac{1}{2}-\epsilon\right)$. Note, that when $t \rightarrow \infty$ and $\mu = const$, then $0 < \lambda < 1/a$. Hence, we have

\begin{align*}
\Pr\left[ X \leq \epsilon t\mu\right] & \leq \exp\left( \lambda \epsilon t\mu -\lambda X_0 - \lambda t \frac{\mu}{2} +\lambda^2 t \log^2 t \right)\\
& = \exp\left( \lambda t \left(\epsilon \mu -\frac{\mu}{2} +\lambda \log^2t\right)  -\lambda X_0\right)\\
& = \exp\left( \lambda t \left(\mu\left(\epsilon  -\frac{1}{2}\right) +\lambda \log^2t\right)  -\lambda X_0\right)\\
& = \exp\left( \lambda t \left(\mu\left(\epsilon  -\frac{1}{2}\right) + \frac{\mu}{2}\left(\frac{1}{2}-\epsilon\right)\right) -\lambda X_0 \right)\\
& = \exp\left( \lambda t \left(\frac{\mu}{2}\left(\epsilon  -\frac{1}{2}\right) \right)-\lambda X_0\right)  \\
& =  \exp\left( -\lambda (t+X_0) \frac{\mu}{2}\left(\frac{1}{2}-\epsilon\right) \right)  \\
& =  \exp\left( - \frac{t+X_0}{4\log^2 t}\mu^2\left(\frac{1}{2}-\epsilon\right)^2 \right).
\end{align*}
And the lemma follows.	
\end{myproof}

\subsection{Contradictory paths and bicycles}
\label{sec:paths}
Unlike $k$-CNFs for larger values of $k$, 2-CNFs have a clear structural feature that indicates whether or not the formula is satisfiable. Let $\phi$ be a 2-CNF on variables $v_1,\dots,v_n$. A sequence of clauses $(l_1,l_2),(\bar l_2,l_3),\dots,(\bar l_{s-1},l_s)$ is said to be a \emph{path} from literal $l_1$ to literal $l_s$. As is easily seen, if there are variables $u,v,w$ in $\phi$ such that there are paths from $u$ to $v$ and $\bar v$, and from $\bar u$ to $w$ and $\bar w$, then $\phi$ is unsatisfiable, see also \cite{ASPVALL1979121}. Such a collection of paths is sometimes called \emph{contradictory paths}. 

On the other hand, if $\phi$ is unsatisfiable, it has to contain a \emph{bicycle}, see~\cite{Chvatal92:mick}. 
A bicycle of length $s$ is a path $(u,l_{1}),(\bar{l}_{1},l_{2}),\dots,(\bar{l}_s,v)$, where the variables associated with literals $l_1,l_2,\dots,l_s$ are distinct, and $u,v \in \{l_{1},\bar{l}_{1},l_{2},\bar{l}_{2},\dots,l_s,\bar{l}_s\}$.

\subsection{The main result}

Now we are ready to state our main result:
\begin{theorem}
Let $\phi \sim \mathbb{C}_{n}^{2}(\xi)$, where $\xi \sim \Class{\alpha}$. Then for $n \rightarrow \infty$
$$
\Pr[\, \phi \text{ is SAT }] = 	\begin{cases}
								0, \text{ when } &  0<\alpha<2,\\
								0, \text{ when } & \alpha = 2 \text{ or } \E\xi^2>3\E\xi,\\
								1, \text{ when}  & \E\xi^2 < 3\E\xi.  
							\end{cases}							
$$
\end{theorem}

If the r.v.\ $\xi$ is distributed according to the zeta distribution, that is, $\Pr\left[\xi = \ell\right]=\frac{\ell^{-\beta}}{\zeta(\beta)}$ for some $\beta > 1$ and where $\zeta(\beta)=\sum_{d \geq 1}d^{-\beta}$ is the Riemann zeta function (note that in this case $\xi\sim\mathcal P(\beta-1)$), then the satisfiabitliy threshold is given by a certain value of $\beta$.

\begin{corollary}\label{cor:main}
Let $\phi \sim \mathbb{C}_{n}^{2}(\xi)$, where the pdf of $\xi$ is $\Pr\left[\xi = \ell\right]=\frac{\ell^{-\beta}}{\zeta(\beta)}$  for some $\beta>1$ and all $\ell \geq 1$. Then there exists $\beta_0$ such that for $n \rightarrow \infty$
$$
\Pr[\, \phi \text{ is SAT }] = 	\begin{cases}
								0, \text{ when } &  1<\beta<\beta_0,\\
								1, \text{ when}  & \beta>\beta_0.  
							\end{cases}							
$$
The value $\beta_0$ is the positive solution of the equation $\E\xi^2= 3\E\xi$, and $\beta_0\approx3.26$.
\end{corollary}

A proof of this theorem constitutes the rest of the paper. We consider each case separately, and the first case is proved in Proposition~\ref{lemma:lower-bound-general-k}, while the other two cases are examined in Propositions~\ref{lemma:3_alpha_alpha0} and~\ref{lemma:alpha0_alpha}.

\section{Satisfiability of $\mathbb{C}_{n}^{2}\left(\xi\right)$, when $\xi \sim \Class{\alpha}$ and $0<\alpha<2$}
\label{sec:sat-infty}

This case is the easiest to analyze. Moreover, we show that the same result holds for any $\phi \sim \mathbb{C}_{n}^{k}\left(\xi\right)$, where $k \geq 2$, when $\alpha<\frac{k}{k-1}$. Hence, the case $0<\alpha<2$ for unsatisfiable 2-CNFs follows. In other words, if $\alpha < \frac{k}{k-1}$, then \textit{any} $k$-CNF formulas from $\mathbb{C}_{n}^{k}\left(\xi\right)$ will be unsatisfiable w.h.p. 

What happens here, is that we expect many variables to have degree $\gg S_n^{(k-1)/k}$. Let us fix $k$ such variables. Then, as it is shown in the proof, the formula $\phi$ contains $(k-1)!\log^kn$ clauses that are formed only from literals of these $k$ variables. However, one of the possible subformulas, which is formed from only $k$ variables, that renders the whole $k$-CNF formula unsatisfiable consists of only $2^k$ clauses.

The next proposition establishes a lower bound of satisfiability threshold for any power-law distributed k-CNF from configuration model.


\begin{proposition}
	\label{lemma:lower-bound-general-k}
	Let $\phi \sim \mathbb{C}_{n}^{k}(\xi)$, where $\xi \sim \Class{\alpha}$, $k \geq 2$ and  $0<\alpha<\frac{k}{k-1}$. Then w.h.p. $\phi$ is unsatisfiable. 
\end{proposition}

\begin{myproof}
	First, recall quantity  $S_n$ that serves as the total number of clones
	$$
	S_n=\sum_{i=1}^{n}\deg(v_i).
	$$
	Since each $\deg(v_i)$ is an independent realization of the r.v. $\xi$ in $\mathbb{C}_{n}^{k}(\xi)$, we have that $\deg(v_i) \overset{d}{=} \xi$.
	
	Next, let's estimate how many variables $v_i$'s in $\phi$ have degrees at least $S_{n}^{(k-1)/k}\log n$:
	\begin{align*}
	\E\left[ \sum_{i=1}^{n} \1{\deg(v_i) \geq S_{n}^{(k-1)/k}\log n} \right] &= \sum_{i=1}^{n} \E\left[\1{\deg(v_i) \geq S_{n}^{(k-1)/k}\log n} \right]\\ 
	& = \sum_{i=1}^{n}\Pr\left[\deg(v_i) \geq S_{n}^{(k-1)/k}\log n\right]\\
	& = \sum_{i=1}^{n}F_{\xi}\left( S_{n}^{(k-1)/k}\log n \right)\\
	& = n F_{\xi}\left( S_{n}^{(k-1)/k}\log n \right).
	\end{align*}
	
	However, since  $\xi \sim \Class{\alpha}$, then $F_{\xi}\left( x \right) \geq W \, x^{-\alpha}$ for some $W>0$ and all $x \geq 1$, hence,
	\begin{align}
	\E\left[ \sum_{i=1}^{n} \1{\deg(v_i) \geq S_{n}^{(k-1)/k}\log n} \right] & = n F_{\xi}\left( S_{n}^{(k-1)/k}\log n \right) \nonumber \\
	& \geq W\, n S_n^{-\alpha(k-1)/k}\log^{-\alpha}n. \label{eq:exp-tmp}
	\end{align}
	Prior to moving to the next steps of the proof, we note that the actual number of variables with degrees at least $S_{n}^{(k-1)/k}\log n$ is distributed according to Binomial distribution, hence, it is concentrated around its mean.

	Next, since $\alpha < \frac{k}{k-1}$, we need to consider 2 cases: first, is when $0<\alpha \leq 1$, and second is for $1 < \alpha < \frac{k}{k-1}$. Thus, for the first case, due to Theorem~\ref{theorem:sum-1}, we have that w.h.p.
	$$
	S_n \leq C\,n^{1/\alpha},
	$$
	where $C>0$ is some constant. Therefore,
	\begin{align*}
	\E\left[ \sum_{i=1}^{n} \1{\deg(v_i) \geq S_{n}^{(k-1)/k}\log n} \right] & \geq W\, n S_n^{-\alpha(k-1)/k}\log^{-\alpha}n\\
	& \geq W\, n \left(C \, n^{1/\alpha}\right)^{-\alpha(k-1)/k}\log^{-\alpha}n\\
	& = \Omega\left( n^{1-(k-1)/k}  \log^{-\alpha}n\right).
	\end{align*}
	
	Hence, we expect polynomially many variables to have large degrees. The same holds in the case, when $1 < \alpha < \frac{k}{k-1}$. Then $S_n=(1+o(1))\,n\E\xi$ w.h.p. due to Theorem~\ref{theorem:sum-2}, and so the expected number of variables with degrees at least $S_n^{(k-1)/k}\log n$ is:
	
	\begin{align*}
	\E\left[ \sum_{i=1}^{n} \1{\deg(v_i) \geq S_{n}^{(k-1)/k}\log n} \right] & \geq W\, n S_n^{-\alpha(k-1)/k}\log^{-\alpha} n\\
	& \geq W\, n \Big((1+o(1)) \, n\E \xi\Big)^{-\alpha(k-1)/k}\log^{-\alpha}n\\
	& = \Omega\left( n^{1-\alpha(k-1)/k}\log^{-\alpha}n \right).
	\end{align*}
	
	And so in both cases we note that we expect many variables having degrees at least $S_n^{(k-1)/k}\log n$. Let us fix $k$ arbitrary variables $v_1,v_2,\dotsc,v_k \in V(\phi)$ such that $\deg(v_1)=\deg(v_2)=\dots=\deg(v_k)=d \geq S_n^{(k-1)/k}\log n$. Next we introduce indicator r.v. $I_c$, which is equal to 1 iff clause $c \in C(\phi)$ consists solely of clones of variables $v_1,v_2,\dotsc,v_k$. Then
	
	$$
	H = \sum_{c \in C(\phi)} I_c
	$$
	is the total number of clauses formed only from clones of variables $v_1,v_2,\dotsc,v_k$.
	
	We show that $H$ is the sum of weakly correlated binary r.vs. and therefore the actual value of $H$ does not deviate much from its expected value. First, the probability that some specific clause $c \in C(\phi)$ is constructed solely from the clones of variables $v_1,v_2,\dots,v_k$ is
	\begin{align*}
	\Pr\left[I_c = 1\right] &=  k! \, \frac{\deg(v_1) }{S_n}\times \frac{\deg(v_2) }{S_n-1}\times \dots \times \frac{\deg(v_k) }{S_n-k+1}\\
	& = k!  \prod_{i=1}^{k}\frac{d-1}{S_n-i+1}\\
	\end{align*}
	
	We also have that  for any two specific clauses $c_0,c_1 \in C(\phi)$
	\begin{align*}
	\Pr\left[ I_{c_1} = 1 \, | \, I_{c_0} = 1 \right] &= k! \, \frac{\deg(v_1) - 1}{S_n-k}\times \frac{\deg(v_2) - 1}{S_n-k-1}\times \dots \times \frac{\deg(v_k) - 1}{S_n-k-i+1}\\
	& = k!  \prod_{i=1}^{k}\frac{d-1}{S_n-k-i+1}\\
	& = (1+o(1)) k! \,  \prod_{i=1}^{k}\frac{d}{S_n-i+1}\\
	& = (1+o(1))\Pr\left[I_{c_1} = 1\right]
	\end{align*}
	Hence, the covariance is
	\begin{align*}
	Cov(I_{c_0}, I_{c_1}) & = \E\left[ I_{c_0}I_{c_1} \right] - \E I_{c_0} \cdot \E I_{c_1}\\
	& = \Pr\left[ I_{c_1} = 1 \, | \, I_{c_0} = 1 \right]\,\Pr\left[I_{c_0} = 1 \right] - \Pr\left[I_{c_0} = 1 \right]\Pr\left[I_{c_1} = 1 \right]\\
	& = (1+o(1))\Pr\left[I_{c_1} = 1\right]\Pr\left[I_{c_0} = 1\right] - \Pr\left[I_{c_0} = 1 \right]\Pr\left[I_{c_1} = 1 \right]\\
	& = o\Big( \Pr\left[I_{c_0} = 1 \right]^2 \Big),
	\end{align*}
	since $ \Pr\left[I_{c_0} = 1 \right]=\Pr\left[I_{c_1} = 1 \right]$. Then the variance of $H$ is at most
	\begin{align*}
	Var[H]=Var\left[\sum_{c \in C(\phi)}I_{c}\right] & = \sum_{c \in C(\phi)}Var[I_c]+\sum\limits_{\substack{c_0 \neq c_1:\\ c_0,c_1 \in C(\phi)}}Cov\left(I_{c_0}, I_{c_1}\right)\\
	& \leq \sum_{c \in C(\phi)}E[I_c]+\sum\limits_{\substack{c_0 \neq c_1:\\ c_0,c_1 \in C(\phi)}}Cov\left(I_{c_0}, I_{c_1}\right)\\
	& = \E\left[\sum_{c \in C(\phi)}I_c\right]+|C(\phi)|^2o\Big( \Pr\left[I_{c^{'}} = 1 \right]^2 \Big), 
	\end{align*}
	where $ c^{'} \in C(\phi)$ is any clause from $\phi$. Thus, we obtain
	\begin{align*}
	Var[H] & \leq \E\left[\sum_{c \in C(\phi)}I_c\right]+|C(\phi)|^2o\Big( \Pr\left[I_{c^{'}} = 1 \right]^2 \Big)\\
	& =  \E\left[H\right]+o\left( \Big( |C(\phi)| \Pr \left[I_{c^{'}} = 1 \right] \Big)^2 \right)\\
	& =  \E\left[H\right]+o\left( \left(\E\left[\sum_{c \in C(\phi)}I_c\right] \right)^2 \right)\\
	& =  \E\left[H\right]+o\left( \E\left[H\right]^2 \right)\\
	& = o\left( \E\left[H\right]^2 \right).
	\end{align*}
	
	Therefore, due to Chebyshev's inequality, it follows that $H$ is concentrated around its expectation, i.e. the expected value $\E \left[H\right]$ serves as a good approximation to the actual value of the r.v. $H$. 
	
	Finally, the expected number of clauses formed only from clones of variables $v_1,v_2,\dotsc,v_k \in V(\phi)$ is
	\begin{align*}
	\E\left[H\right] & = \E\left[\sum_{c \in C(\phi)}I_c\right] = \sum_{c \in C(\phi)}\Pr\left[I_c=1\right]\\
	& = \sum_{c \in C(\phi)}k!\,\frac{\deg(v_1)}{S_n}\times\frac{\deg(v_2)}{S_n-1}\times\dots\times\frac{\deg(v_k)}{S_n-k+1}\\
	& =  (1+o(1))\,\sum_{c \in C(\phi)}k!\, \left(\frac{d}{S_n}\right)^k\\
	& =  (1+o(1))\,|C(\phi)|k!\, \left(\frac{d}{S_n}\right)^k\\
	& =  (1+o(1))\,\frac{S_n}{k}k!\, \left(\frac{d}{S_n}\right)^k, \text{ since } |C(\phi)|=\frac{S_n}{k}\\
	& =  (1+o(1))\,(k-1)!\, \frac{d^k}{S_n^{k-1}}\\
	& \geq  (1+o(1))\,(k-1)!\, \frac{\left(S_n^{(k-1)/k}\log n\right)^k}{S_n^{k-1}}, \text{ since } d \geq S_n^{(k-1)/k}\log n\\
	& = (1+o(1))\,(k-1)!\, \log^k n.
	\end{align*}
	
	Hence, since $H = (1+o(1))\E \left[H\right]$ w.h.p., and $\E \left[H\right] \geq (1+o(1))\,(k-1)!\, \log^k n$, it follows that the number of clauses formed solely of clones of the fixed $k$ variables grows together with $n$. However, as it was pointed earlier, we need only a $2^k$ clauses subformula to make $\phi$ unsatisfiable. Thus, $\phi$ is \textit{UNSAT} w.h.p. when $0 < \alpha < \frac{k}{k-1}$. 
\end{myproof}

After proving the above proposition, result for 2-CNF from $\mathbb{C}_n^2(\xi)$ naturally follows. 

\begin{corollary}
	\label{corollary:2_alpha_3}
	Let $\phi \sim \mathbb{C}_{n}^{2}(\xi)$, where $\xi \sim \Class{\alpha}$, such that $0<\alpha<2$. Then w.h.p. $\phi$ is unsatisfiable. 
\end{corollary}

\section{Satisfiability of $\mathbb{C}_{n}^{2}\left(\xi\right)$, when $\xi \sim \Class{\alpha}$ and $\alpha=2$ or $\E\xi^2>3\E\xi$}
\label{sec:sat-tspan}

\subsection{The inequality $\E\xi^2>3\E\xi$}

Analysis of this and subsequent cases mainly follows the approach suggested in~\cite{Cooper2007}, where they deal with random 2-SAT instances having prescribed literal degrees. In other words, the assumption in \cite{Cooper2007} is that the degree sequences $d^+_1,\dots,d^+_n$ and $d^-_1,\dots,d^-_n$ are fixed, and a random 2-CNF is generated as in the configuration model. Then two quantities play a very important role. The first one is the sum of all degrees $S_n=\sum_{i=1}^n(d^+_i+d^-_i)$ (we use our notation) and the second one is the number of pairs of complementary clones $T_n=\sum_{i=1}^n d^+_id^-_i$. It is then proved that a 2-CNF with a given degree sequence is satisfiable w.h.p.\ if and only if $2T_n<(1-\varepsilon)S_n$ for some $\varepsilon>0$. We will quickly show that the conditions $\alpha=2$ and $\E\xi^2>3\E\xi$ imply the inequality $2T_n>(1+\varepsilon)S_n$ w.h.p., see Lemma~\ref{lem:TS-condition}, and therefore a random 2-CNF in this case should be unsatisfiable w.h.p. The problem however is that Cooper et al.\ only prove their result under a significant restrictions on the maximal degree of literals, $\Delta<n^{1/11}$. By Lemma~\ref{lemma:delta} the maximal degree of literals in our case tends to be much higher, and we cannot directly utilize the result from \cite{Cooper2007}. Therefore we follow the main steps of the argument in \cite{Cooper2007} changing parameters, calculations, and in a number of cases giving a completely new proofs. 

\begin{lemma}\label{lem:TS-condition}
Let $\phi \sim \mathbb{C}_{n}^{2}(\xi)$, where $\xi \sim \Class{\alpha}$ and $\alpha=2$ or $\E\xi^2>3\E\xi$. Let also 
$
S_n=\sum_{i=1}^n d_i \quad \text{and} \quad 
T_n=\sum_{i=1}^n d^+_id^-_i.
$
Then w.h.p.\ $2T_n>(1+\varepsilon)S_n$.
\end{lemma}

\begin{myproof}
Let us first consider the case, when $\alpha >2$ and $\E\xi^2>3\E\xi$. Then by Lemma~\ref{lemma:moments} and Theorem~\ref{theorem:sum-2}, we have that w.h.p.
$$
S_n=\sum_{i=1}^{n}d_{i}=(1+o(1))\sum_{i=1}^{n}\E d_i = (1+o(1))\, n\E\xi,
$$
since $d_i \overset{d}{=}\xi$. Likewise, since $\alpha > 2$, we also have that w.h.p.
$$
T_n=\sum_{i=1}^{n}d_{i}^{+}d_{i}^{-}=(1+o(1))\sum_{i=1}^{n}\E \left[d_{i}^{+}d_{i}^{-}\right] = (1+o(1))\, n\frac{\E\xi^2-\E\xi}{4},
$$
where the last equality follows from Lemma~\ref{lemma:dd}.

Hence, when $\E\xi^2>3\E\xi$, we have that w.h.p.
\begin{align*}
\frac{2T_{n}}{S_{n}} = (1\pm o(1))\frac{\E\xi^2-\E\xi}{2\E\xi}= (1\pm o(1))\left( \frac{\E\xi^2}{2\E\xi}-\frac{1}{2} \right)>1.
\end{align*}

Now we consider the case $\alpha=2$. Unfortunately, then $\E \left[d_{i}^{+}d_{i}^{-}\right] = \infty$ for any $i \in [1 \dots n]$, and so we cannot claim that $T_n$ is concentrated around its mean. Nevertheless, the quantity $\frac{2T_n}{S_n}$ is still greater than 1 in this case.

Since $\xi\sim\Class{2}$, there are constants $V,W$ such that 
$
W\,\ell^{-2}\le F_{\xi}(\ell)\le V\,\ell^{-2}.
$
We construct auxiliary random variables $\xi_{\varepsilon} \sim \Class{2 + \varepsilon}$ for $\varepsilon>0$. Later we will argue that $\xi_{\varepsilon}$ can be chosen such that $\E\xi_{\varepsilon}^{2} > 3\E\xi_{\varepsilon}$.  Specifically, let $\xi_{\varepsilon}$ be such that $F_{\xi_{\varepsilon}}(1)=1$ and $F_{\xi_{\varepsilon}}(\ell)=W\,\ell^{-2-\varepsilon}$ for $\ell>1$. 

Let $T_n^{\varepsilon}$ be the number of pairs of complementary clones in formula $\phi_0 \sim \mathbb{C}_{n}^{2}(\xi_{\varepsilon})$. Since $\Pr\left[\xi_{\varepsilon} \geq \ell\right] \leq \Pr\left[\xi \geq \ell\right]$ for any $\ell \geq 1$, we have that
\begin{equation}
\Pr\left[2T_n > S_n\right]  
\geq \Pr\left[2T_n^{\varepsilon} > S_n\right],  \label{eq:probability_tn}
\end{equation}
due to the stochastic dominance of the r.v. $T_n$ over $T_n^\varepsilon$. As is easily seen, for sufficiently small $\varepsilon$ we have $\E\xi_{\varepsilon}^{2} > 3\E\xi_{\varepsilon}$. Therefore, by the first part of the proof $2T_n^{\varepsilon} > S_n$ w.h.p. The result follows.

Thus, in either case we obtain that for some $\mu>0$ w.h.p. 
$
\frac{2T_{n}}{S_{n}} = 1 + \mu.\label{2Tn_Sn}
$
\end{myproof}

In what follows, we will always assume that $\alpha>2$.

\subsection{TSPAN}

The process of generating a random 2-CNF in the configuration model can be viewed as follows. After creating a pool of clones, we assign each clone a polarity, making it a clone of a positive or negative literal. Then we choose a random partioning of the set of clones into 2-element sets. The important point here is that in the process of selection of a random matching we pair clones up one after another, and it does not matter in which order a clone to match is selected, as long as it is paired with a random unpaired clone.

Our goal is to show that our random 2-CNF $\phi$ contains contradictory paths. In order to achieve this we exploit the property above as follows. Starting from a random literal $p$ we will grow a set $span(p)$ of literals reachable from $p$ in the sense of paths introduced in Section~\ref{sec:paths}. This is done by trying to iteratively extend $span(p)$ by pairing one of the unpaired clones of the negation of a literal from $span(p)$. The details of the process will be described later. The hope is that at some point $span(p)$ contains a pair of literals of the form $v,\bar v$, and therefore $\phi$ contains a part of the required contradictory paths. To obtain the remaining part we run the same process starting from $\bar p$.

To show that this approach works we need to prove three key facts:\\[1mm]
-- that $span(p)$ grows to a certain size with reasonable probability (Lemma~\ref{lemma:s2}),\\[1mm]
-- that if $span(p)$ has grown to the required size, it contains a pair $v,\bar v$ w.h.p.\ (Lemma~\ref{lemma:bad_event}), and \\[1mm]
-- that the processes initiated at $p$ and $\bar p$ do not interact too much w.h.p.\ (Lemma~\ref{lemma:span}).\\[2mm]
\indent
Since the probability that $span(p)$ grows to the required size is not very high, most likely this process will have to be repeated multiple times. It is therefore important that the probabilities above are estimated when some clones are already paired up, and that all the quantities involved are carefully chosen.

\smallskip

We now fill in some details. The basic ``growing'' algorithm is \textproc{TSPAN} (short for \textit{truncated span}), see Algorithm~\ref{alternative-algorithm}. Take a literal and pick a clone $p$ associated with it. Then partition the set $\mathcal{S}$ of all clones into 3 subsets: the set $\mathcal{L}(p)$ of ``live'' clones from which we can grow the span, the set $\mathcal{C}$ of paired (or ``connected'') clones, and the set $\mathcal{U}$ of ``untouched'' yet clones. We start with $\mathcal{L}(p)=\{p\}$, $\mathcal{U} = \mathcal{S}-\{p\}$, and empty set $\mathcal{C}$. 

\textproc{TSPAN} works as follows: while the set of live clones is not empty, pick u.a.r.\ clone $c_1$ from the live set, and pair it u.a.r. with any non-paired clone $c_2 \in \mathcal{U} \cup \mathcal{L}(p) \setminus \{c_1\}$. Since clones $c_1$ and $c_2$ are paired now, we move them into the set of paired clones $\mathcal{C}$, while removing them from both sets $\mathcal{L}(p)$ and $\mathcal{U}$ to preserve the property that the sets $\mathcal{C},\mathcal{U}$, and $\mathcal{L}(p)$ form a partition of $\mathcal{S}$.

Next, we identify the literal $l$ which clone $c_2$ is associated with, and we move all the complementary clones of $\bar{l}$ from the set of untouched clones $\mathcal{U}$ into $\mathcal{L}(p)$. The idea of this step is, when we add an edge $(c_1,c_2)$, where $c_2$ is one of the $l$'s clones, to grow the span further we will need to add another directed edge $(c_3,\,\cdot\,)$, where $c_3$ is one of the clones belonging to $\bar{l}$. Hence, we make all clones of $\bar{l}$ live, making them available to pick as a starting point during next iterations of \textproc{TSPAN}. This way we can grow a span, starting from the clone $p$, and then the set
$$
span(p) = \{ c \in \mathcal{S} \, | \, c \text{ is reachable from  } p \},
$$ 
contains all the clones, which are reachable from the clone $p$ (or literal that is associated with $p$) at a certain iteration of \textproc{TSPAN}. We call this set a \emph{$p$-span}.

The version of \textproc{TSPAN} given in Algorithm~\ref{alternative-algorithm} takes as input sets $\mathcal{C},\mathcal{L},\mathcal{U}$ (which therefore do not have to be empty in the beginning of execution of the procedure), a maximal number of iterations $\tau$, and a maximal size of the set of live clones. It starts by using the given sets, $\mathcal{C},\mathcal{L},\mathcal{U}$, stops after at most $\tau$ iterations or when $\mathcal L$ reaches size $\sigma$.

\begin{algorithm}
\caption{Procedure TSPAN}
\label{alternative-algorithm}
\begin{algorithmic}[1]
\Procedure{TSPAN}{$\mathcal{C},\mathcal{L},\mathcal{U}, \sigma,\tau$}
   \While {$0 < |\mathcal{L}| \leq \sigma$ \textbf{and} less than $\tau$ pairings performed}
    	\State Pick u.a.r. a live clone $c_1 \in \mathcal{L}$
    	\State Pick u.a.r. an unpaired clone $c_2 \in \mathcal{U} \cup \mathcal{L} \setminus \{c_1\}$
    	\State Pair clones $c_1$ and $c_2$, i.e.
    	\State $\quad \mathcal{C} \leftarrow \mathcal{C} \cup \{c_1,\,c_2\}$
    	\State $\quad \mathcal{L} \leftarrow \mathcal{L} \setminus \{c_1,\,c_2\}$ 
    	\State $\quad \mathcal{U} \leftarrow \mathcal{U} \setminus \{c_1,\,c_2\}$
    	\State Let $w$ be the literal associated with $c_2$ 
    	\State Make live the clones associated with $\bar{w}$, i.e 
    	\State Let 
    	\State $\quad \kappa(\bar{w}) = \{ c \in \mathcal{S} \, | \, c $ is associated with $ \bar{w} \}$ \label{alg:step5}
    	\State $\quad L \leftarrow L \cup (U \cap \kappa(\bar{w}))$
    	\State $\quad U \leftarrow U \setminus \kappa(\bar{w})$  \label{alg:step6}
    \EndWhile
\EndProcedure
	\end{algorithmic}
\end{algorithm}

\subsection{Searching for contradictory paths}

The procedure \textproc{TSPAN} is used to find contradictory paths as follows:\\[2mm]
{\sc Step 1.} 
Pick a variable and a pair of its complementary clones $p,q$.\\[1mm]
{\sc Step 2.} 
Run \textproc{TSPAN} starting from $p$ for at most $s_1=\mysone$ steps. If $\mathcal{L}(p)$ becomes empty during the process, or if $q$ gets included into $span(p)$, or if in the end $|\mathcal{L}(p)|<\sigma=s_1\mu/6$ ($\mu$ is determined by the value $2T_n/S_n$, see Lemma~\ref{lemma:expectation}), declare failure.\\[1mm]
{\sc Step 3.}
Run \textproc{TSPAN} starting from $q$ and the current set $\mathcal C$ of paired clones for at most $s_1=\mysone$ steps. If $\mathcal{L}(q)$ becomes empty during the process, or if $|\mathcal{L}(q)\cap\mathcal{L}(q)|=\Theta(s_1)$, or if in the end $|\mathcal{L}(q)|<\sigma$, declare failure.\\[1mm]
{\sc Step 4.}
Run \textproc{TSPAN} starting from $\mathcal L(p)$ and the current set $\mathcal C$ of paired clones for at most $s_2=\mystwo$ steps. If $\mathcal{L}(p)$ becomes empty during the process, declare failure.\\[1mm]
{\sc Step 5.}
Similarly, run \textproc{TSPAN} starting from $\mathcal L(q)$ and the current set $\mathcal C$ of paired clones for at most $s_2=\mystwo$ steps. If $\mathcal{L}(q)$ becomes empty during the process, declare failure.\\[3mm]
\indent
If a failure is declared at any step, we abandon the current pair $p,q$ and pick another variable and a pair of its complementary clones keeping the current set $\mathcal C $ of paired clones that will be used in the next round. Also, even if all the Steps are successful, but the constructed span does not contain contradictory paths, we also declare a failure. It is important that the set $\mathcal C $ never grows too large, that is, it remains of size $|\mathcal C|=o(n)$. This implies that the number of restarts does not exceed $K=\myk$.

\smallskip

The next lemma shows how we exploit the value of $\mu$, since it acts as an approximation to the number of newly added live  into the set of live clones when $\frac{2T_{n}}{S_{n}}>1$. However, first, we need to introduce several variables. Let $\mathcal{L}_i$, $\mathcal{U}_i$, and $\mathcal{C}_i$ are the live, untouched, and connected sets respectively after the $i$-th iteration of some execution of \textproc{TSPAN}.  Additionally we have 
$
L_i = \left| \mathcal{L}_i \right|,\ 
C_i = \left| \mathcal{C}_i \right|,\ 
U_i = \left| \mathcal{U}_i \right|.
$
Also let $X_{i}$ indicate the change in the number of live clones after performing the \textit{i}th iteration, i.e.
$
X_{i}=L_{i}-L_{i-1}.
$

\begin{lemma}
\label{lemma:expectation}
Let $\frac{2T_{n}}{S_{n}} = 1 + \mu$, where $\mu > 0$. Then for any $t\leq |\mathcal{C}| = o\left(n\right)$, we have
$$
\E \left[X_{t} \, | \, X_1,\dots,X_{t-1}\right] \geq \mu/2.
$$
\end{lemma}  

Before proving Lemma~\ref{lemma:expectation} we show a couple of useful auxiliary results regarding $X_i$'s and $L_i$'s:

\begin{lemma}
\label{lemma:right-tail}
There exists constant $V > 0$, such that for any $d \geq 1$ and $i \leq |\mathcal{C}|=o\left(n\right)$
\begin{align}
\Pr\left[ X_i \geq d \, | \, X_1, X_2, \cdots, X_{i-1} \right] \leq V\,d^{1-\alpha}. \label{eq:right-tail}
\end{align}

\end{lemma}
\begin{myproof}
	Notice that $X_i \geq 0$ indicates that at the $i$-th iteration, the \textproc{TSPAN} picked a clone $c_i$ to pair with from the set of untouched clones $\mathcal{U}_{i-1}$. Assume that $c_i$ is associated with literal $l_i$. Then $X_i = \deg(\bar{l}) - 1$ (minus one comes from the fact that according to the \textproc{TSPAN} algorithm, $\mathcal{L}_{i-1}$ always loses one clone).  
	
	Hence, for the live set to gain $d$ clones, $c_i$ must be one of the clones associated with literals whose complements have exactly $d+1$ clones. Thus,
	\begin{align}
	\Pr\left[X_i \geq d \, | \, X_1,X_2,\cdots, X_{i-1}\right] & \leq  \frac{1}{S_n-2i-1}\E\left[ \sum\limits_{l \in L(\phi)}\deg(l)\cdot\1{\deg(\bar{l}) \geq d+1} \right] \nonumber  \\
	& =  \frac{1}{S_n-2i-1} \sum\limits_{l \in L(\phi)}\E\left[ \deg(l)\cdot\1{\deg(\bar{l}) \geq d+1} \right]. \label{eq:tmp-4}
	\end{align}
	In the above estimate of the probability we claim that the total number of clones of the literals with degree at least $d+1$ is well-approximated by its mean. To see why this is true, note that the tail function of the r.v. $\deg(l)\cdot\1{\deg(\bar{l}) \geq d+1} $ is upper bounded by the right tail function of the r.v. $\deg(l)=d^{+}=d^{-}$. Hence, for $\alpha>2$, we expect that
	$$
	\sum\limits_{l \in L(\phi)} \deg(l)\cdot\1{\deg(\bar{l}) \geq d+1} = (1+o(1))\,\sum\limits_{l \in L(\phi)}\E\left[ \deg(l)\cdot\1{\deg(\bar{l}) \geq d+1} \right],
	$$
	due to Theorem~\ref{theorem:sum-2}.
	
	Now, let us fix some specific literal $l$ and let $v$ be its corresponding variable. Then
	\begin{align*}
	\E\left[ \deg(l)\cdot\1{\deg(\bar{l}) \geq d+1} \right] & = \sum_{\ell=d+1}^{\infty}\sum_{\ell_0 = d+1}^{\ell}(\ell-\ell_0) \, \Pr\left[\deg(\bar{l})=\ell_0 \, | \deg(v) = \ell\right]\Pr\left[\deg(v) = \ell \right]\\
	& \leq \sum_{\ell=d+1}^{\infty}\sum_{\ell_0 = d+1}^{\ell}\ell \, \Pr\left[\deg(\bar{l})=\ell_0 \, | \deg(v) = \ell\right]\Pr\left[\deg(v) = \ell \right]\\
	& \leq \sum_{\ell=d+1}^{\infty}\sum_{\ell_0 = d+1}^{\ell}\ell \, \Pr\left[Bin\left(\deg(v), \frac{1}{2}\right)=\ell_0 \, | \deg(v) = \ell\right]\Pr\left[\deg(v) = \ell \right]\\
	& = \sum_{\ell=d+1}^{\infty}\sum_{\ell_0 = d+1}^{\ell}\ell \, \binom{\ell}{\ell_0} \frac{1}{2^\ell}\Pr\left[\deg(v) = \ell \right]\\
	& = \sum_{\ell=d+1}^{\infty}\frac{\ell}{2^\ell}\Pr\left[\deg(v) = \ell \right]\sum_{\ell_0 = d+1}^{\ell}\binom{\ell}{\ell_0} \\
	& \leq \sum_{\ell=d+1}^{\infty}\frac{\ell}{2^\ell}\Pr\left[\deg(v) = \ell \right]\sum_{\ell_0 = 0}^{\ell}\binom{\ell}{\ell_0} \\
	& =  \sum_{\ell=d+1}^{\infty}\frac{\ell}{2^\ell}\Pr\left[\deg(v) = \ell \right] \, 2^\ell \\
	& \leq \sum_{\ell=d}^{\infty}\ell\Pr\left[\deg(v) = \ell \right].
	\end{align*}
	
	Next, we apply summation by parts
	\begin{align*}
	\E\left[ \deg(l)\cdot\1{\deg(\bar{l}) \geq d+1} \right] & \leq \sum_{\ell=d}^{\infty}\ell\Pr\left[\deg(v) = \ell \right]\\
	& = \sum_{\ell=d}^{\infty}\ell\Big(\Pr\left[\deg(v) \geq \ell \right]-\Pr\left[\deg(v) \geq \ell +1\right]\Big)\\
	& = \sum_{\ell=d}^{\infty}\ell\,\Pr\left[\deg(v) \geq \ell \right]-\sum_{\ell=d+1}^{\infty}(\ell-1)\,\Pr\left[\deg(v) \geq \ell\right]\\
	& = d\,\Pr\left[\deg(v) \geq d \right]+\sum_{\ell=d+1}^{\infty}\ell\,\Pr\left[\deg(v) \geq \ell \right]-\\
	& \qquad \qquad - \sum_{\ell=d+2}^{\infty}(\ell-1)\,\Pr\left[\deg(v) \geq \ell\right]\\
	& = d\,\Pr\left[\deg(v) \geq \ell \right]+\sum_{\ell=d+1}^{\infty}\Pr\left[\deg(v) \geq \ell \right].
	\end{align*}
	
	Recall that $\deg(v) \overset{d}{=}\xi \sim \Class{\alpha}$, thus, we obtain that for a fixed $l$
	\begin{align*}
	\E\left[ \deg(l)\cdot\1{\deg(\bar{l}) \geq d+1} \right] & \leq d\,\Pr\left[\deg(v) \geq d \right]+\sum_{\ell=d+1}^{\infty}\Pr\left[\deg(v) \geq \ell \right]\\
	& = d\,\Pr\left[\xi \geq d \right]+\sum_{\ell=d+1}^{\infty}\Pr\left[\xi \geq \ell \right]\\
	& = d\,F_{\xi}(d)+\sum_{\ell=d+1}^{\infty}F_{\xi}(\ell)\\
	& \leq V\,d^{1-\alpha}+V\sum_{\ell=d+1}^{\infty}\ell^{-\alpha}\\
	& \leq V_r\, d^{1-\alpha},
	\end{align*}
	for some constant $V_r>0$.
	
	Therefore,~\eqref{eq:tmp-4} can be further simplified
	\begin{align*}
	\Pr\left[X_i \geq d \, | \, X_1,X_2,\cdots, X_{i-1}\right] & \leq  \frac{1}{S_n-2i-1} \sum\limits_{l \in L(\phi)}\E\left[ \deg(l)\cdot\1{\deg(\bar{l}) \geq d+1} \right] \nonumber \\
	& \leq \frac{1}{S_n-2i-1} \sum\limits_{l \in L(\phi)}V_r\, d^{1-\alpha} \nonumber \\
	& \leq \frac{2n}{S_n-2i-1} V_r\, d^{1-\alpha},
	\end{align*}
	since  $|L(\phi)| \leq 2n$. Recall that w.h.p. $S_n=(1+o(1))\, n\E\xi$. Then for $i=o\left(n\right)$ we obtain
	\begin{align*}
	\Pr\left[X_i \geq d \, | \, X_1,X_2,\cdots, X_{i-1}\right] & \leq \frac{2n}{S_n-2i-1} V_r\, d^{1-\alpha}\\
	& = \frac{2n}{(1+o(1))\,n\E\xi-2i-1} V_r\, d^{1-\alpha}\\
	& \leq 2V_r \, d^{1-\alpha},
	\end{align*}
	since $\E\xi \geq 1$. Finally, denote by $V:= 2V_r$,
	and the lemma follows with
	$$
	\Pr\left[ X_i \geq d \, | \, X_1, X_2, \cdots, X_{i-1} \right] \leq V\,d^{1-\alpha}. 
	$$
\end{myproof}

Next, we  show that $L_i = o(n)$ as long as $i$ is not too large. However, instead of showing this result directly, we will prove a somewhat more general statement from which the desired property naturally follows. 

\begin{lemma}
\label{lemma:max-size}
Let the sequence $\{\xi_i\}_{i=1}^{cn}$, where $c>0$ is constant, contains $cn$ independent copies of the r.v. $\xi$ which has the right tail for any $\ell \geq 1$
$$
F_{\xi}(\ell)=\Pr\left[ \xi \geq \ell \right] \leq V\,\ell^{-\alpha},
$$
with $V>0$ and $\alpha > 1$. Let $A \subset \{\xi_i\}_{i=1}^{cn}$ be any subset of size $t = n^{\beta} \; (0< \beta < 1)$. Then w.h.p. 
$$
\sum_{X \in A}X = O\left( n^{\beta + \frac{1-\beta}{\alpha}  }  \right).
$$
\end{lemma}

\begin{myproof}
Let us call a variable $\xi_j$ ``heavy'', when $\xi_j \geq  n^{\frac{1-\beta}{\alpha}}$. Then the expected sum of ``heavy'' variables in the original sequence $\{\xi_i\}_{i=1}^{cn}$ is
\begin{align*}
\E \left[\sum_{1 \leq i \leq cn}\xi_i\,\1{\xi_i \geq n^{\frac{1-\beta}{\alpha}}} \right] & =  cn\sum_{d \geq n^{\frac{1-\beta}{\alpha}}}  d\Pr\left[ \xi=d \right]\\
&=cn \sum_{d \geq n^{\frac{1-\beta}{\alpha}}}  d\Big(\Pr\left[ \xi \geq d \right] - \Pr\left[ \xi \geq d + 1 \right]\Big)\\
&= cn \sum_{d \geq n^{\frac{1-\beta}{\alpha}}}  d\Pr\left[ \xi \geq d \right] - cn\sum_{d \geq n^{\frac{1-\beta}{\alpha}}+1}  (d-1)\Pr\left[ \xi \geq d \right]\\
& = cn^{1+\frac{1-\beta}{\alpha}}\Pr\left[\xi \geq n^{\frac{1-\beta}{\alpha}}\right] + cn\sum_{d \geq n^{\frac{1-\beta}{\alpha}}+1}\Pr\left[ \xi \geq d \right]\\
& = cn^{1+\frac{1-\beta}{\alpha}} F_{\xi}\left(n^{\frac{1-\beta}{\alpha}}\right)+ cn\sum_{d \geq n^{\frac{1-\beta}{\alpha}}+1}F_{\xi}\left(d\right).
\end{align*}
Next, recall that $F_{\xi}\left(n^{\frac{1-\beta}{\alpha}}\right) \leq V\,n^{\beta-1}$. Then
\begin{align*}
\E \left[\sum_{1 \leq i \leq n}\xi_i\,\1{\xi_i \geq n^{\frac{1-\beta}{\alpha}}} \right] & \leq cn^{1+\frac{1-\beta}{\alpha}} F_{\xi}\left(n^{\frac{1-\beta}{\alpha}}\right)+ cn\sum_{d \geq n^{\frac{1-\beta}{\alpha}}+1}F_{\xi}\left(d\right)\\
& \leq cVn^{\beta + \frac{1-\beta}{\alpha}} + cVn\sum_{d \geq n^{\frac{1-\beta}{\alpha}}+1}d^{-\alpha}\\
& = cVn^{\beta + \frac{1-\beta}{\alpha}}  + O\left( n^{1 + (1-\alpha)\frac{1-\beta}{\alpha}  }  \right)\\
& = O\left( n^{\beta + \frac{1-\beta}{\alpha}  }  \right).
\end{align*}

Since the right tail functions of the r.vs. $\xi_i\,\1{\xi_i \geq n^{\frac{1-\beta}{\alpha}}}$ is upper bounded by the right tail function of the original non-truncated variable $\xi$, the sum of ``heavy'' variables is concentrated around its mean. Therefore, even if $A$ contained only the ``heavy'' variables, the resulting sum would be at most $O\left( n^{\beta + \frac{1-\beta}{\alpha}  }  \right)$. However, if we pick ``non-heavy'' variables, meaning the ones that have degree at most $n^{\frac{1-\beta}{\alpha}}$, then again the sum would be at most $O\left( n^{\beta + \frac{1-\beta}{\alpha}  }  \right)$. And so no matter what variables the set $A$ includes, the sum of its elements is $O\left( n^{\beta + \frac{1-\beta}{\alpha}  }  \right)$ w.h.p.	
\end{myproof}

However, our concern was to bound the size of the live set $L_i$. And the next corollary shows that we expect $L_i=o(n)$ as long as $i$ is not too large.

\begin{corollary}
\label{cor:live-set-max-size}
When $t \leq |\mathcal{C}| = o\left(n\right)$, then $L_t =\sum_{i=1}^{t}X_i = o(n)$ holds w.h.p. 
\end{corollary}

\begin{myproof}
	Proof is pretty straightforward now, given Lemma~\ref{lemma:max-size}. We have at most $2n$ literals (it is not exactly $2n$, since some variables may produce only one literal). And the right tail function of the literals is 
	$$
	\Pr\left[ \deg(l) \geq \ell \right] \leq V\,\ell^{-\alpha},
	$$
	due to~\eqref{eq:literal-tail}. Since $\alpha > 2$, and after applying Lemma~\ref{lemma:max-size} we obtain that
	$$
	L_t = \sum_{i=1}^{t}X_i = o(n)
	$$
	holds w.h.p.
\end{myproof}


Now we are in a position to prove Lemma~\ref{lemma:expectation}.

\begin{myproof}[of Lemma~\ref{lemma:expectation}]
Assume that at the \textit{t}-th iteration of \textproc{TSPAN} procedure we paired clones $(p,p')$. Clearly, according to the algorithm, $p \in \mathcal{L}_{t-1}$, thus, $\mathcal{L}_i$ loses at least one clone, and, with probability $\frac{L_{t-1}}{S_{n}-2t+1}$, $p'$ can be also from $\mathcal{L}_{t-1}$, making the live set to lose another clone.

Next let us call variable $v_{i}$ \textit{undisturbed}, if none of its clones were paired or made live. Then, if $p$ is paired with a positive clone of the undisturbed variable $v_{i}$ (and we can do so in $d_{i}^{+}$ ways with uniform probability $\frac{1}{S_{n}-2t-1}$), then the live set will gain $d_{i}^{-}$ new clones. Similarly, when $p$ is paired with a negative clone of $v_{i}$ (which can be done in $d_{i}^{-}$ ways with uniform probability $\frac{1}{S_{n}-2t-1}$), then the live set gains $d_{i}^{+}$ new clones.   

Note, it may happen that $p$ is paired with clones of a ``disturbed'' variable, which may add some clones into $\mathcal{L}_t$, but since we are looking for a lower bound, we can ignore this case. Therefore,
 
\begin{align*}
\E \left[X_{t} \, | \, X_1,\dots,X_{t-1}\right] & = \E\left[L_{t} - L_{t-1}\, | \, X_1,\dots,X_{t-1}\right] \\
& \geq -1 - \frac{L_{t-1}}{S-2t-1} + \frac{1}{S-2t-1}\sum_{\text{j undisturbed}}\left(d_{j}^{+}d_{j}^{-} + d_{j}^{-}d_{j}^{+}\right) \\
& \geq -1 + \frac{2}{S_{n}-2t-1}\Big( \sum_{j=1}^{n}d_{j}^{+}d_{j}^{-} - \sum_{\text{j disturbed}}d_{j}^{+}d_{j}^{-} - \frac{L_{t-1}}{2}\Big).
\end{align*}
Recall that $\sum_{j=1}^{n}d_{j}^{+}d_{j}^{-}=T_{n}$, and w.h.p. 
$$
T_n=(1+o(1))\,n\frac{\E\xi^2-\E\xi}{4}.
$$
Then 
\begin{align*}
\E \left[X_{t} \, | \, X_1,\dots,X_{t-1}\right] 
& \geq -1 + \frac{2}{S_{n}-2t-1}\Big( \sum_{j=1}^{n}d_{j}^{+}d_{j}^{-} - \sum_{\text{j disturbed}}d_{j}^{+}d_{j}^{-} - \frac{L_{t-1}}{2}\Big)\\
& \geq -1 + \frac{2}{S_{n}-2t-1}\Big( T_n - \sum_{\text{j disturbed}}d_{j}^{+}d_{j}^{-} - \frac{L_{t-1}}{2}\Big).
\end{align*}

The sum $\sum\limits_{\text{j disturbed}}d_{j}^{+}d_{j}^{-}$ is the sum of at most $t=o\left(n\right)$ random variables $d_{j}^{+}d_{j}^{-}$. Note, though, that these variables $d_{j}^{+}d_{j}^{-}$ are not independent, nor identically distributed in the aforementioned sum, since during the first iterations the \textproc{TSPAN} procedure favours  products of complementary literals having larger numbers of clones.

To bound this sum, let us introduce the set $A$ of \textit{disturbed} variables. Then
$$
\sum\limits_{\text{j disturbed}}d_{j}^{+}d_{j}^{-} = \sum_{v_j \in A}d_{j}^{+}d_{j}^{-}.
$$
We know that $|A| \leq t = o\left(n\right)$, and $A \subset T$, where $T=\{ d_i^{+}d_{i}^{-} \, | \, v_i \in V(\phi) \}$
is the set of all products of degrees of complementary literals. Now, every element  $d^{+}d^{-} \in T$ has the right tail function $\Pr\left[d^{+}d^{-}  \geq \ell\right] \leq 2V\,\ell^{-\alpha/2}$~\eqref{eq:pair-tail} for some $V>0$ and every $\ell \geq 1$. Thus, since $\alpha > 2$, and according to Lemma~\ref{lemma:max-size}, we obtain that 
$$
\sum_{v_i \in A}d_{j}^{+}d_{j}^{-} = o(n),
$$
which in turn means that w.h.p. $\sum\limits_{\text{j disturbed}}d_{j}^{+}d_{j}^{-} = o(n)$. As for the $L_{t-1}$ term, we know from Corollary~\ref{cor:live-set-max-size} that it remains $o(n)$ as well. 

Hence,   
\begin{align*}
\E \left[X_{t} \, | \, X_1,\dots,X_{t-1}\right] & \geq -1 + \frac{2}{S_{n}-2t-1}\Big( T_n- \sum_{\text{j disturbed}}d_{j}^{+}d_{j}^{-} - \frac{L_{t-1}}{2}\Big)\\
& \geq -1 + \frac{2}{S_{n}-o(n)}\Big( T_{n} - o(n)\Big) \\
& \geq -1 + \frac{2T_{n}}{S_{n}}\Big( 1 - o(1)\Big) \\
& \geq -1 + (1+\mu)( 1 - o(1)) \\
& \geq \mu/2.
\end{align*}
\end{myproof}

Next, we bound the probability of failure in each of {Steps 2--5}. We start with {\sc Step 2} assuming that the number of paired clones is $o(n)$.

\begin{lemma}[{\sc Step 2}]
\label{lemma:span}
(1) Let $s_1=\mysone$. If \textproc{TSPAN} starts with a live set containing only a single point $L_0=1$, time bound $\tau=s_1$, the live set size bound $\sigma=s_1\mu/6$, and the number of already paired clones $|\mathcal{C}|=o\left(n\right)$, then with probability at least $\frac{1}{2s_1}$ \textproc{TSPAN} terminates with the live set of size at least $\sigma$. \\[1mm]
(2) For any fixed clone $q$, the probability it will be paired in $s_{1}=\mysone \leq t=o\left(n\right)$ steps of the algorithm, is at most $o\left(\frac{1}{s_1}\right)$. 
\end{lemma}

\begin{myproof}
(1) The \textproc{TSPAN} procedure may terminate at the moment $i < \tau$ due to one of two reasons: first, when $L_{i}$ hits 0, and second, when $L_{i} = \sigma$. To simplify analysis of the lemma, instead of dealing with conditional probabilities that the live set hasn't paired all its clones, we suggest to use a slightly modified version of \textproc{TSPAN}, which \textit{always} runs for $\tau$ steps. 

The modified version works exactly as the original \textproc{TSPAN} procedure when the live set has at least one clone. But if at some moment, the live set has no clones to pick, we perform a ``restart'': we restore the sets $\mathcal{L}, \mathcal{C}$, and $\mathcal{D}$ to the states they'd been before the first iteration of \textproc{TSPAN} procedure occurred. After that we continue the normal pairing process. Although during restarts we reset the values of the sets, the counter that tracks the number of iterations the \textproc{TSPAN} has performed is never reset, and keeps increasing with every iteration until the procedure has performed pairings $\tau$ times, or the live set was able to grow up to size $\sigma$, and only then the \textproc{TSPAN} terminates.   

Now, let $r_{i}=1$ represents a ``successfull'' restart that started at $i$ iteration, meaning during this restart the live set accumulated $\sigma$ clones, while $r_{i}=0$ means there was no restart or the live set became empty. What we are looking for $\Pr[r_{1}=1]$, since this probability is identical to the probability that the original \textproc{TSPAN} was able to grow the live set to the desired size. Next, we can have at most $\tau$ restarts, and, since the very first restart has the most time and we expect the live set to grow in the long run, it follows that it stochastically dominates over other $r_{i}$'s. Thus,
\begin{align*}
\Pr\left[L_{s_1} \geq s_{1}\mu/6\right] \leq \Pr\left[\sum_{i=1}^{s_{1}}r_{i} \geq 1\right] \leq \E\sum_{i=1}^{s_{1}}r_{i} \leq s_{1}\E r_{1}=s_1\Pr[r_{1}=1]
\end{align*}
from which we obtain the probability that the \textproc{TSPAN} terminates with large enough live set from the very first try: 
\begin{align}
P := \Pr[r_{1}=1] \geq \frac{\Pr[L_{s_{1}} \geq s_{1}\mu/6]}{s_{1}}. \label{eq:p}
\end{align}

Now what is left is to obtain bounds on the right-hand side probability. We have a random process
$$
L_{s_{1}}=\sum_{i=1}^{s_{1}}\left(L_{i}-L_{i-1}\right)=\sum_{i=1}^{s_{1}}X_{i},
$$ 
which consists of steps $X_i$, each having the right tail (Lemma~\ref{lemma:right-tail})
$$
\Pr\left[X_i \geq \ell \, | \, X_1,\dots,X_{i-1}\right] \leq V\,\ell^{-\alpha},
$$
and positive expectation (Lemma~\ref{lemma:expectation})
$$
\E\left[X_i  \, | \, X_1,\dots,X_{i-1}\right]  \geq \frac{\mu}{2}.
$$	

Therefore, according to \textit{Azuma-like inequality} (Lemma \ref{lemma:azuma}), we obtain that
\begin{align*}
\Pr\left[L_{s_{1}} \leq s_{1}\mu/6\right] = \Pr\left[ L_{s_{1}} \leq \left(s_{1}\frac{\mu}{2}\right)\frac{1}{3} \right] \leq \exp\left( - \frac{s_1}{4\log^2 s_1}\frac{\mu^2}{576} \right)
\end{align*}
Fixing $s_{1}=\mysone$, we have for some constant $C>0$
$$
\Pr\left[L_{s_{1}} \leq s_{1}\mu/6 \right] \leq \exp\left(-C\frac{\mysone}{\log^2 n}\right)  = o(1) \leq 1/2.
$$
Thus, from~\eqref{eq:p} follows $P \geq \frac{\Pr[L_{s_{1}} \geq s_{1}\mu/6 ]}{s_{1}}=\frac{1}{2s_{1}}$, which proves the first part of the lemma.

\smallskip

(2) To pair clone $q$, we must select it uniformly among $|\mathcal{S}|-|\mathcal{C}|$ non-paired clones. Hence, due to Union bound, we have
\begin{align*}
\Pr\left[q \notin \mathcal{U}_{s_1} \right] & \leq \Pr\left[ \bigcup_{i=1}^{s_1}\{ q \in \mathcal{C}_{i} \} \right]\\
& \leq \sum_{i=1}^{s_1}\Pr\Big[ q \in \mathcal{C}_{i} \Big]\\
& = s_1 \frac{1}{|\mathcal{S}|-|\mathcal{C}|}\\
& = \frac{s_1}{ S_n - o(n) }, \text{ since } |\mathcal{C}| = t = o(n), \text{ and } |\mathcal{S}| = S_n\\
& = (1+o(1)) \, \frac{\mysone}{n\E\xi }, \text{ since } S_n= (1+o(1))\,n\E\xi \text{ w.h.p.}\\
& = O\left( n^{\frac{ \alpha+4 }{6(\alpha+1)}-1}\right)\\
& = O\left( n^{-\frac{ 1 }{3}}\right)\\
& = o\left(\frac{1}{s_1}\right). 
\end{align*}
\end{myproof}

Note that in Lemma~\ref{lemma:span}(1) the size of $\mathcal L(p)$ can be slightly greater than $\sigma$, as it may increase by more than 1 in the last iteration. Also, in  Lemma~\ref{lemma:span}(2) the bound on the probability is only useful when $s_1$ is sufficiently large.

The probability that both runs of \textproc{TSPAN} for $p$ and $q$ are successful is given by the following 

\begin{lemma}[{\sc Step 3}]
\label{lemma:s1}
The probability that two specific clones $p$ and $q$ accumulate $s_{1}\mu/7$ clones in their corresponding live sets $\mathcal{L}$ during the execution of {\sc Steps 2,3}, 
such that the span from clone $p$ doesn't include $q$ nor make it live, is at least $\frac{1}{5s_1^2}$. 
\end{lemma}

We start with an auxiliary lemma.

\begin{lemma}
	The \textproc{TSPAN} procedure will pair at most $o(s_1)$ clones from the set of live clones $\mathcal{L}(p)$, while constructing the span from $q$, when $|\mathcal{C}| = o(n)$.
\end{lemma}
\begin{myproof}
	Since the \textproc{TSPAN} from $p$ and $q$ runs for at most $s_1$ steps, and $|\mathcal{L}(p)|=o(n)$ w.h.p. due to Corollary~\ref{cor:live-set-max-size}, we have that the expected number of paired clones, which belonged to the set $\mathcal{L}(p)$ is at most
	\begin{align*}
		\E\left[ \sum_{i=1}^{s_1}\1{\text{u.a.r. picked clone } c \in \mathcal{L}(p)} \right] &=\sum_{i=1}^{s_1}\Pr\left[\text{u.a.r. picked clone } c \in \mathcal{L}(p)\right] \\ 
		&=\sum_{i=1}^{s_1}\frac{|\mathcal{L}(p)|}{ |\mathcal{S}| - |\mathcal{C}| } \\
		&=\sum_{i=1}^{s_1}\frac{o(n)}{ (1+o(1))\,n\E\xi - o(n) }, 
	\end{align*}
	where the last step follows from $|\mathcal{S}| = S_n = (1+o(1))\,n\E\xi$ w.h.p. Hence,
	\begin{align*}
	\E\left[ \sum_{i=1}^{s_1}\1{\text{u.a.r. picked clone } c \in \mathcal{L}(p)} \right] &=\sum_{i=1}^{s_1}\frac{o(n)}{ (1+o(1))\,n\E\xi - o(n) }  \\
	& = (1+o(1)) s_1 \frac{o(n)}{n}\\
	& = o(s_1).
	\end{align*}
	
	Moreover, since the process $\sum_{i=1}^{s_1}\1{\text{u.a.r. picked clone } c \in \mathcal{L}(p)}$ forms a binomial trial, it follows that the actual number of paired clones does not deviate much from its expectation, and hence, w.h.p.
	$$
	\sum_{i=1}^{s_1}\1{\text{u.a.r. picked clone } c \in \mathcal{L}(p)} = o(s_1). 
	$$  	
\end{myproof}

Thus, after growing span from clone $p$, there is a good probability that $q$ isn't paired. So we can run the same \textproc{TSPAN} algorithm for clone $q$ now, and Lemma~\ref{lemma:span} suggests that we will be able to construct the second span of size $s_{1}\mu/6$ with probability at least $\frac{1}{2s_1}$.

However, we need to make sure that while the TSPAN process constructs span from $q$, it doesn't pair too many clones that were marked as ``live'' and placed into $\mathcal{L}(p)$. But since $L_i=o(n)$ w.h.p. for $i \leq t = o(n)$, this ''bad`` event is unlikely to happen.

\begin{lemma}
The probability that two specific clones $p$ and $q$ accumulate $s_{1}\mu/7$ clones in their corresponding live sets $\mathcal{L}$ during the execution of \textproc{TSPAN} algorithm, such that the span from clone $p$ doesn't pair $q$ nor make it ``live'', is at least $\frac{1}{5s_1^2}$. 
\end{lemma}
\begin{myproof}
Let $t_0 \leq t = o\left(n\right)$ be the time when we picked complementary clones $p$ and $q$ as roots for growing spans, and let $L_{t_0+s_1}(p)$ and $L_{t_0+2s_1}(q)$ be the sizes of the live sets $\mathcal{L}(p)$ and $\mathcal{L}(q)$ respectively after performing $s_1$ iterations of \textproc{TSPAN} first for clone $p$ and then for $q$. Then
\begin{align*}
& \Pr\Big[\{ L_{t_0+s_1}(p) \geq  s_1\mu/6  \} \wedge \{ L_{t_0+2s_1}(q) \geq s_1\mu/6  \} \wedge \{ q \in \mathcal{U}_{t+s_1} \} \Big] = \\
& = \Pr\Big[ L_{t_0+s_1}(p) \geq s_1\mu/6  \, | \,  \{ L_{t_0+2s_1}(q) \geq s_1\mu/6  \} \wedge \{ q \in \mathcal{U}_{t+s_1} \} \Big] \times \\
& \qquad \qquad \times \Pr\Big[\{ L_{t_0+2s_1}(q) \geq s_1\mu/6  \} \wedge \{ q \in \mathcal{U}_{t+s_1} \}\Big]\\
& \geq  \Pr\Big[ L_{t_0+s_1}(p) \geq s_1\mu/6  \, | \,  \{ L_{t_0+2s_1}(q) \geq s_1\mu/6  \} \wedge \{ q \in \mathcal{U}_{t+s_1} \} \Big] \times \\
& \qquad \qquad \times \Big( \Pr[ L_{t_0+2s_1}(q) \geq s_1\mu/6] - \Pr[  q \notin \mathcal{U}_{t+s_1}] \Big) \\
& \geq \frac{1}{2s_{1}} \Big( \frac{1}{2s_{1}} - \Pr\left[  q \notin \mathcal{U}_{t+s_1}\right] \Big), 
\end{align*}
since $\Pr\left[ L_{t_0+s_1}(p) \geq s_1\mu/6\right] =\Pr\left[ L_{t_0+2s_1}(q) \geq s_1\mu/6\right]  > \frac{1}{2s_1}$  due to Lemma~\ref{lemma:span}. Next recall that 
$$
\Pr\left[  q \notin \mathcal{U}_{t+s_1}\right]=o\left(\frac{1}{s_1}\right)
$$ 
from Lemma~\ref{lemma:span}.2, and so it follows that
\begin{align*}
\Pr\Big[\{ L_{t_0+s_1}(p) \geq s_1\mu/6 \} \wedge \{ L_{t_0+2s_1}(q) \geq s_1\mu/6  \} \wedge \{ q \in \mathcal{U}_{t+s_1} \} \Big]
& \geq \frac{1}{2s_{1}} \Big( \frac{1}{2s_{1}} - \Pr[  q \notin \mathcal{U}_{t+s_1}] \Big)\\
& \geq \frac{1}{2s_{1}} \Big( \frac{1}{2s_{1}} - o\left(\frac{1}{s_1}\right)\Big) \\
& \geq \frac{1}{5s_{1}^{2}}, 
\end{align*}
and result of the lemma follows.
\end{myproof}

Next, we show that we can grow the spans for another $s_2$ steps, while keeping  sizes of the respective live sets of order at least $s_1\mu/8$.

\begin{lemma}[{\sc Steps 4,5}]
\label{lemma:s2}
Assume that $p$- and $q$-spans were both able to accumulate at least $s_1\mu/8$ live clones after $s_1=\mysone$ steps, and $q$ is not in the $p$-span. Then with probability $ 1-o(1)$, \textproc{TSPAN} will be able to perform another $s_{2}=\mystwo$ iterations, and $L_{s_1+j} \geq s_{1}\mu/8$ for every $0 \leq j \leq s_{2}$ for each clone $p$ and $q$.
\end{lemma}

We start by showing that \textproc{TSPAN} in {\sc Step 4} does not decimate the $\mathcal L(q)$.

\begin{lemma}
	Let $t_0 \leq t = |\mathcal{C}|=o(n)$ be the moment of time, when we picked complementary clones $p$ and $q$, and let $L_{t_0+s_1}(p) \geq s_1\mu/7$ and  $L_{t_0+2s_1}(q) \geq s_1\mu/7$ be the sizes of the live sets of clones $p$ and $q$ respectively after growing each for $s_1$ steps. Then the number of clones from $\mathcal{L}(q)$ that will be paired during $s_2$ iterations of the  \textproc{TSPAN} procedure, while expanding the $p$-span, is at most $o(s_1)$ w.h.p. 
\end{lemma}

\begin{myproof}
 The expected number of clones, which were marked as ``live'' by $q$-span, while \textproc{TSPAN} is expanding the $p$-span is
	\begin{align*}
	\E\left[ \sum_{i=1}^{s_2} \1{\text{u.a.r. picked clone } c \in \mathcal{L}(q) } \right] & = \sum_{i=1}^{s_2} \Pr\left[ \text{u.a.r. picked clone } c \in \mathcal{L}(q)  \right]\\
	& = \sum_{i=1}^{s_2} \frac{|\mathcal{L}(q)|}{|\mathcal{S}|-|\mathcal{C}|}\\
	& = (1+o(1))\, \sum_{i=1}^{s_2} \frac{|\mathcal{L}(q)|}{n\E\xi},
	\end{align*}
	since $|\mathcal{S}|-|\mathcal{C}|=S_n-o(n)=(1+o(1))\,n\E\xi$ w.h.p. Now recall that $|\mathcal{L}(q)|$ is the sum of at most $s_1$ r.vs. $\{X_{t_0+s_1+i}\}_{i=1}^{s_1}$, where each $X_j$ represents the number of clones of the literals being added into $\mathcal{L}(q)$. Hence, $\{X_{t_0+s_1+i}\}_{i=1}^{s_1} \subset D$, where the set $D$ is the multi-set of degrees of literals in $\phi$, i.e. $D:= \{ \deg(l) \, | \,  l \in L(\phi) \}$ and $|D| \leq 2n$. Moreover, due to Corollary~\ref{cor:literal-tail}, each element in $D$ is a r.v. with the right tail function $\Pr\left[\deg(l) \geq \ell \right] \leq V\, \ell^{-\alpha}$.
	
Then, according to Lemma~\ref{lemma:max-size}, we have that w.h.p. 
$$
|\mathcal{L}(q)| = O\left(  n^{ \frac{\alpha+4}{6(\alpha+1)} + \frac{1 - \frac{\alpha+4}{6(\alpha+1)}}{\alpha} }  \right) = O\left( n^{\frac{\alpha^2+9\alpha+2}{6\alpha(\alpha+1)}} \right).
$$ 
Thus, we obtain
	\begin{align*}
	\E\left[ \sum_{i=1}^{s_2} \1{\text{u.a.r. picked clone } c \in \mathcal{L}(q) } \right] & = (1+o(1))\, \sum_{i=1}^{s_2} \frac{|\mathcal{L}(q)|}{n\E\xi}\\
	& = (1+o(1))\, s_2 \frac{|\mathcal{L}(q)|}{n\E\xi}\\
	& = (1+o(1))\, s_2 \, O\left( \frac{n^{\frac{\alpha^2+9\alpha+2}{6\alpha(\alpha+1)}}}{n} \right)\\
	& = O\left( \mystwo \times  n^{\frac{\alpha^2+9\alpha+2}{6\alpha(\alpha+1)}-1} \right)\\
	& = O\left( n^{\frac{\alpha^2+9\alpha+2}{12\alpha(\alpha+1)}} \right)\\
	& = o\left(\mysone\right), \text{ when } \alpha>2\\
	& = o(s_1).
	\end{align*}
	And since $\sum_{i=1}^{s_2} \1{\text{u.a.r. picked clone } c \in \mathcal{L}(q) } $ forms a binomial trial, it follows that the actual number of clones from $\mathcal{L}(q)$ that get paired by the $p$-span is concentrated around its expectation. Hence, we do not expect more than $o(s_1)$ clones from $\mathcal{L}(q)$ to be paired, while expanding the span from $p$.
\end{myproof}

Finally, we are in a position to prove Lemma~\ref{lemma:s2}

\begin{myproof}
When we start from a live set with size of order at least $s_{1}\mu/8$, to be able to grow the span for another $s_{2}$ steps, we need to make sure that the size of the live set never drops to zero. First, recall that $Ks_{1}+s_{2} =  o\left(n\right)$, thus, according to Lemma~\ref{lemma:expectation}, for any $j = o\left(n\right)$
$$
\E \left[ X_{s_1+j} \, | \, X_1,\dots, X_{s_1+j-1} \right] \geq \mu/2>0
$$ 
and for any $d \geq 1$ 
$$
\Pr\left[ X_{s_1+j} \geq d \, | \, X_1,\dots, X_{s_1+j-1}  \right] \leq V\,d^{-\alpha}.
$$
Then, after applying the Azuma-like inequality (\ref{lemma:azuma}), we obtain
\begin{align*}
  \Pr\left[L_{s_1+j} \leq s_{1}\frac{\mu}{8}\right] & \leq \Pr\left[L_{s_1+j} \leq \left(s_{1}\frac{\mu}{2}\right)\frac{1}{3}\right] \\
  & = \Pr\left[ L_{s_1} + \sum_{i=1}^{j}X_{s_1+i} \leq \left(s_{1}\frac{\mu}{2}\right)\frac{1}{3} \right]\\
& \leq \exp\left( - \frac{j+L_{s_1}}{4\log^2 s_2}\frac{\mu^2}{576} \right)\\
& \leq \exp\left( - \frac{j+s_1 \mu }{\log^2 s_2}\frac{\mu^2}{4032} \right), \text{ since } L_{s_1} \geq s_1\mu/7\\
& \leq \exp\left( - \frac{s_1 \mu }{\log^2 s_2}\frac{\mu^2}{4032} \right).
\end{align*}  

Recall that $s_1=\mysone$ and $s_2=\mystwo$. Hence, we obtain that for  \textit{specific} $1 \leq j \leq s_2$
$$
\Pr\left[L_{s_1+j} \leq s_{1}\frac{\mu}{8}\right] \leq \exp\left( - \frac{s_1 \mu }{\log^2 s_2}\frac{\mu^2}{4032} \right) \leq \exp\left(-C\frac{\mysone}{\log^2 n}\right),
$$
for some constant $C>0$. Then the probability that the live set will drop below the $s_1\mu/7$ level during \textit{any} of the $1 \leq j \leq s_{2}$ steps is, by \textit{Union bound},
\begin{align*}
\Pr[\bigcup\limits_{j=1}^{s_{2}}\{L_{s_1+j} \leq s_{1}\frac{\mu}{8}\}] & \leq \sum_{j=1}^{s_{2}}\Pr\left[L_{s_1+j} \leq s_{1}\frac{\mu}{8}\right] \\
& \leq \sum_{j=1}^{s_{2}}\exp\left(-C\frac{\mysone}{\log^2 n}\right) \\
& = s_2\cdot \exp\left(-C\frac{\mysone}{\log^2 n}\right) \\
& = o(1). 
\end{align*} 
\end{myproof}

Therefore, after the \textproc{TSPAN} finishes constructing the $p$-span, the size of the ``live'' set of $q$-span is at least
$$
|\mathcal{L}(q)| \geq L_{t_0+2s_1} - o(s1) \geq s_1\frac{\mu}{7} - o(s_1) \geq  s_1\frac{\mu}{8}.
$$
To keep our calculations as simple as possible, we will assume that $|\mathcal{L}(p)|\geq s_1\frac{\mu}{8}$ as well, since we have proved that if \textproc{TSPAN} succeeds at the first stage for both $p$- and $q$-spans, then w.h.p. $|\mathcal{L}(p)|\geq s_1\frac{\mu}{7}$.

Finally, we show that w.h.p. the spans produced in {\sc Steps 1--5}, provided no failure occurred, contain contradictory paths. In other words, we are looking for the probability that spans do not contain complement clones after growing them for $s_1+s_{2}$ steps, i.e. at each step \textproc{TSPAN} was choosing only untouched clones from the set $\mathcal{U}$.

\begin{lemma}[Contradictory paths]
\label{lemma:bad_event}
If for a pair of complementary clones $p,q$ {\sc Steps 1--5} are completed successfully, the probability that $span(p)$ or $span(q)$ contains no 2 complementary clones is less than $\exp\Big(-n^{\frac{\alpha^2-\alpha-2}{12\alpha(\alpha+1)}} \Big)$.
\end{lemma}

\begin{myproof}
Let $B$ (``bad'') be the event that after performing $s_1+s_{2}$ steps, there were no two complementary clones ever paired. This means, as was mentioned previously, that at each step \textproc{TSPAN} was choosing u.a.r. only untouched clones $c \in \mathcal{U}$ among $\mathcal{L}\cup \mathcal{U}$ clones. 

Recall that sets $\mathcal{U},\mathcal{L},$ and $\mathcal{C}$ form a partition of the set $\mathcal{S}$ of all clones. Then at any moment of time $i$ we have $S_{n}=C_{i}+U_{i}+L_{i}=2i+U_{i}+L_{i}$. Hence,
\begin{align*}
\Pr[B]  & = \prod_{i=1}^{s_1+s_{2}}\frac{U_{i}}{L_{i}+U_{i}-1} \\
		& = \prod_{i=1}^{s_1+s_{2}}\Big(1 - \frac{L_{i}}{L_{i}+U_{i}-1}\Big) \\
		& = \prod_{i=1}^{s_1+s_{2}}\Big(1 - \frac{L_{i}}{S_{n}-2i-1}\Big) \\
		& \leq \prod_{i=0}^{s_{2}}\Big(1 - \frac{L_{s_1+i}}{S_{n}}\Big).
\end{align*}
Next, Lemma~\ref{lemma:s2} implies, that $L_{s_1+i} \geq s_{1}\mu/8$ for $0 \leq i \leq s_{2}$, so
\begin{align*}
\Pr[B]  & = \prod_{i=0}^{s_{2}}\left(1 - \frac{L_{t+i}}{S_{n}}\right) \\
		& \leq \prod_{i=0}^{s_{2}}\left(1 - \frac{s_{1}\mu}{8S_{n}}\right) \\
		& \leq \exp\Big(-\frac{s_{1}s_{2}\mu}{8S_{n}}\Big).
\end{align*}
Having $s_{1}=\mysone$, $s_{2}=\mystwo$, and $S_{n}=(1+o(1))\,n\E\xi$ w.h.p., we obtain
\begin{align*}
\Pr[B]  & \leq \exp\Big(-\frac{s_{1}s_{2}\mu}{8S_{n}}\Big) \\
		& = \exp\Big(- \frac{\mu \, \mysone  \mystwo}{8(1+o(1))\,n\E\xi }\Big) \\
		& \leq \exp\Big(-n^{\frac{\alpha^2-\alpha-2}{12\alpha(\alpha+1)}} \Big), 
\end{align*}
which proves the lemma.
\end{myproof}

This completes the proof in the case $\alpha=2$ or $\E\xi^2>3\E\xi$, and the next proposition summarizes the result.

\begin{proposition}
\label{lemma:3_alpha_alpha0}
Let $\phi \sim \mathbb{C}_{n}^{2}(\xi)$, where $\xi \sim \Class{\alpha}$ and $\alpha=2$ or $\E\xi^2>3\E\xi$. Then w.h.p. $\phi$ is unsatisfiable.
\end{proposition}

\begin{myproof}
Recall that our algorithm grows spans for a sequence of pairs of complementary clones waiting for the first success, that is, contradictory paths. First, we estimate the probability of finding contradictory paths for some pair of complementary clones in this sequence.  Suppose that at some point we picked clones $p$ and $q$, and let $\{ L_{s_1}(p) \geq s_1\mu/7 \}$, $\{ L_{2s_1}(q) \geq s_1\mu/7 \}$, and $\{ q \in \mathcal{U}_{s_1}  \}$ be the events from Lemma~\ref{lemma:s1}. Then by Lemma~\ref{lemma:s1} with probability at least $\frac{1}{5s_{1}^{2}}$ we will be able to accumulate in each of the live sets at least $s_{1}\mu/7$ elements, while constructing spans from the complementary clones $p$ and $q$. Thus,
$$
\Pr[A_1] :=\Pr\Big[\{ L_{s_1}(p) \geq s_1\mu/7  \} \wedge \{ L_{2s_1}(q) \geq s_1\mu/7  \} \wedge \{ q \in \mathcal{U}_{s_1} \} \Big] \geq \frac{1}{5s_{1}^{2}}.
$$

Next, Lemma~\ref{lemma:s2} implies that if event $A_1$ happens, then with probability at least $1-o(1)$ we will be able to grow both spans for another $s_2$ iterations, such that  sizes of the live sets never drop below $s_1\mu/8$ clones, i.e.
\begin{align*}
\Pr[A_2 \, | \, A_1] & :=  \Pr\left[ \bigcap_{j=1}^{s_2}\{ L_{s_1+j}(p) \geq s_1\mu/8 \} \wedge \bigcap_{j=1}^{s_2}\{ L_{2s_1+s_2+j}(q) \geq s_1\mu/8 \}  \, | \, A_1 \right]\\
&  \geq 1-o(1).
\end{align*} 

If the events $A_1$ and $A_2$ happen, then by Lemma~\ref{lemma:bad_event}  with probability at least $ 1 - o(1)$ the corresponding span will contain 2 complementary clones. Let us denote by $Comp(p)$ the event that there exist 2 complementary clones in the span originating at $p$.

Then, the probability of event $Comp(p,q)$ that a pair of \textit{fixed} complementary clones $p$ and $q$, after completing {\sc Steps 1--5} form contradictory paths is at least 
\begin{align*}
\Pr\left[ Comp(p,q) \right] & = \Pr[Comp(p)\wedge Comp(q) \wedge A_1 \wedge A_2] \\
& \geq \Pr\left[Comp(p)\wedge Comp(q)\right]\,\Pr\left[A_1 \wedge A_2\right]\\ 
& = \Pr[\bar{B}]^2\, \Pr\left[A_2 \,|\, A_1\right]\Pr\left[A_1\right]\\
 & \geq \left(1-\exp\Big(-n^{\frac{\alpha^2-\alpha-2}{12\alpha(\alpha+1)}} \Big)\right)^2 \left(1-o(1)\right) \frac{1}{5s_{1}^{2}} \\
 &= (1-o(1))\frac{1}{5s_{1}^{2}} 
 \ \  \geq \frac{1}{6s_{1}^{2}} \\
 & =: P_{final},
\end{align*}
when $n \rightarrow \infty$. In other words, $P_{final}$ is a lower bound of the probability that 2 \textit{specific} complementary clones $p$ and $q$ will form contradictory paths. However, recall that we have  $T_{n}=\sum_{i=1}^{n}d_i^{+}d_i^{-}$ different pairs of complementary clones, and, moreover, w.h.p. $T_n=(1+o(1))\, n\frac{\E\xi^2-\E\xi}{4}$, when $\alpha>2$. Since we repeat the \textproc{TSPAN} procedure for at most $K= \myk \ll T_n$ different pair of clones, the probability that none of the picked pairs form a contradiction path is at most 
\begin{align*}
\Pr\left[ \bigcap_{i=1}^{K}\overline{Comp(p_i,q_i)} \right] & \leq \prod_{i=1}^{K}\left(1-\Pr\left[Comp(p,q) \right]\right) \\
& = \prod_{i=1}^{K}\left(1-\frac{1}{6s_1^2}\right) 
\ \  \leq \exp\left(-\frac{K}{6s_1^2}\right)\\
& =\exp\left(-\frac{1}{6} \myk \, n^{-2\frac{\alpha+4}{6(\alpha+1)}} \right), \text{ since } s_1=\mysone\\
& =\exp\left(-\frac{1}{6}n^{\frac{\alpha-2}{4(\alpha+1)}}\right)
\ \  = o(1), \text{ since } \alpha>2,
\end{align*}
when $n \rightarrow \infty$. Thus, we've obtained that when $\E\xi^2 > 3\E\xi$, then w.h.p. we expect that some pair of clones will form contradictory paths, witnessing unsatisfiability of the formula $\phi$. And so the proposition follows.
\end{myproof}

\section{Satisfiability of $\mathbb{C}_{n}^{2}\left(\xi\right)$, when $\xi \sim \Class{\alpha}$ and $\E\xi^2<3\E\xi$}
\label{sec:sat-paths}

Chv\'atal and Reed~\cite{Chvatal92:mick} argue that if \textit{2-SAT} formula $\phi$ is unsatisfiable, then it contains a \textit{bicycle}, see Section~\ref{sec:paths}. Thus, the absence of bicycles may serve as a convenient witness of formula's satisfiability. The general idea of this section is to show that w.h.p. there are no bicycles in $\phi \sim \mathbb{C}_{n}^{2}(\xi)$, when $\xi \sim \Class{\alpha}$ and $\E\xi^2<3\E\xi$. 

Intuitively, when $\E\xi^2<3\E\xi$, then we expect $\frac{2T_{n}}{S_{n}}=1-\mu' > 0$, where $\mu'>0$ is some small number. As it was shown in Lemma~\ref{lemma:expectation}, the latter quantity approximates the number of newly added live clones, when running the \textproc{TSPAN} procedure. Since \textproc{TSPAN} always performs at least one iteration of growing the span, it may add at most $\Delta$ clones into the live set after constructing the very first span from the root. After that each subsequent iteration adds \textit{on average} $\approx\frac{2T_{n}}{S_{n}}$ new live clones. So after running the \textproc{TSPAN} for $j$ iterations, where $j \rightarrow \infty$, when $n \rightarrow \infty$, then 
we expect the live set to contain around
$$
L_{t^*}=\Delta\left(\frac{2T_{n}}{S_{n}}\right)^{j}=\Delta(1-\mu')^{j} \leq \Delta e^{-j\mu'}
$$
clones. Therefore, after $O(\log n)$ iterations, the live set becomes empty, and \textproc{TSPAN} terminates. Thus, we expect paths of length at most $O(\log n)$, which is not enough for bicycles to occur.

More formally we first show that in the case $\frac{2T_{n}}{S_{n}}=1-\mu'$ a random formula is unlikely to contain long paths.

\begin{lemma}\label{lem:short-paths}
If $\frac{2T_{n}}{S_{n}}=1-\mu'<1$, then paths in $\phi$ are of length $O\left(\log n\right)$, w.h.p.
\end{lemma}

\begin{myproof}
Let $k_{0}=\lceil \frac{6}{\mu'}\log n \rceil$ and let $P_{k}$ be the number of paths of length $k$ in $G_{I}(\phi)$. Also recall that by $L(\phi)$ we denote the set of all literals in $\phi$, while $\deg(l)$, where $l$ is some literal, denotes the total number of clones of the literal $l$. Then
\begin{align*}
\E P_{k_{0}} & \leq \sum_{l_1,\dots,l_{k_{0}} \in L(\phi)}\Pr\left[l_1 \implies l_2 \implies \dots \implies l_{k_{0}} \in G_{I}(\phi)\right] \\
			& \leq \sum_{l_1,\dots,l_{k_{0}} \in L(\phi)} \frac{2\deg(\bar{l}_{1})\deg(l_2)}{S_{n}-1}\frac{2\deg(\bar{l}_{2})\deg(l_3)}{S_{n}-3}\dots\frac{2\deg(\bar{l}_{k_{0}-1})\deg(l_{k_{0}})}{S_{n}-2k_{0}+3} \\
			& \leq \sum_{l_1,l_{k_{0}}}\frac{2\deg(\bar{l}_{1})\deg(l_{k_{0}})}{S_{n}-2k_{0}} \sum_{l_2,\dots,l_{k_{0}-1}}\prod_{i=2}^{k_{0}-1}\frac{2\deg(\bar{l}_{i})\deg(l_{i})}{S_{n}-2k_{0}} \\
			& \leq \sum_{l_1,l_{k_{0}}}\frac{2\Delta^{2}}{S_{n}-2k_{0}} \sum_{l_2,\dots,l_{k_{0}-1}}\prod_{i=2}^{k_{0}-1}\frac{2\deg(\bar{l}_{i})\deg(l_{i})}{S_{n}-2k_{0}} \\
			& \leq \frac{2n^{2}\Delta^{2}}{S_{n}-2k_{0}} \Big( \frac{2T_{n}}{S_{n}-2k_{0}} \Big) ^{k_{0}-2} \\
			& \leq (1+o(1))\frac{2n^{2}\Delta^{2}}{S_{n}} \Big( (1+o(1))\frac{2T_{n}}{S_{n}} \Big) ^{k_{0}-2}.
\end{align*}
Since w.h.p. $S_{n}=(1+o(1))\, n\E\xi$, and $\frac{2T_{n}}{S_{n}}=1-\mu'$, we have
\begin{align*}
\E P_{k_{0}} & \leq (1+o(1))\frac{2n^{2}\Delta^{2}}{S_{n}} \Big( (1+o(1))\frac{2T_{n}}{S_{n}} \Big) ^{k_{0}-2} \\
			 & \leq \frac{2n\Delta^{2}}{\E\xi}\Big((1+o(1))(1-\mu')\Big)^{k_{0}-2}\\
			 & \leq 2n\Delta^{2}\Big((1+o(1))(1-\mu')\Big)^{k_{0}-2},
\end{align*}
where the last inequality follows from the fact that $\E\xi \geq 1$ for any r.v. $\xi \geq 1$. Next, given large enough $n$, we can assume that $(1+o(1))(1-\mu') \leq 1-\frac{\mu'}{2}$. Hence,
\begin{align*}
\E P_{k_{0}} & \leq 2n\Delta^{2}\Big((1+o(1))(1-\mu')\Big)^{k_{0}-2}\\
			 & \leq 2n\Delta^{2}\Big(1-\frac{\mu'}{2}\Big)^{k_{0}-2} \\
			 & \leq Cn\Delta^{2}\Big(1-\frac{\mu'}{2}\Big)^{k_{0}}, \text{ where } C>0 \text{ is some constant} \\
			 & \leq Cn\Delta^{2}e^{-\mu'k_{0}/2}\\
			 & \leq Cn\Delta^{2}e^{-3\log n} , \text{ since } k_0= \lceil \frac{6}{\mu'}\log n \rceil\\
			 & \leq C\frac{\Delta^{2}}{n^2}.
\end{align*}
Next, since $\alpha>2$, we have that $\Delta = o\left(n^{1/2}\right)$ (see Lemma~\ref{lemma:delta}). However, then
\begin{align*}
\E P_{k_{0}} & \leq C\frac{\Delta^{2}}{n^2} =C\frac{o(n)}{n^2}= o(1),
\end{align*}
and so w.h.p. there are no paths of length greater than $k_{0}$ in $G_{I}(\phi)$.
\end{myproof}

Next we give a straightforward estimation of the number of `short' bicycles.

\begin{lemma}\label{lem:short-bicycles1}
If $\frac{2T_{n}}{S_{n}}=1-\mu'<1$, then for any $k$ the expected number of bicycles of length $k$ is at most $(1+o(1))\frac{2\Delta^{2}}{S_{n}}\left((1+o(1))(1-\mu')\right)^{k}$.
\end{lemma}

\begin{myproof}
Let $B_{k}$ be the number of bicycles of length $k$. Then simple calculation verifies that
\begin{align*}
\E B_k & \leq  \sum_{\substack{ l_1,\dots,l_{k} \in L(\phi) \\ u,v \in \{l_1,\bar{l}_1,\dots, l_k,\bar{l}_k\}  }}
\frac{2\deg(\bar{l}_{1})\deg(l_2)}{S_{n}-1}\dots\frac{2\deg(\bar{l}_{k-1})\deg(l_{k})}{S_{n}-2k+5}\frac{\deg(u)\deg(l_1)}{S_{n}-2k+3} \frac{\deg(v)\deg(\bar{l}_{k})}{S_{n}-2k+1} \\
& \leq \sum_{\substack{ l_1,\dots,l_{k} \in L(\phi) \\ u,v \in \{l_1,\bar{l}_1,\dots, l_k,\bar{l}_k\}  }}
\frac{2\deg(\bar{l}_{1})\deg(l_2)}{S_{n}-1}\dots\frac{2\deg(\bar{l}_{k-1})\deg(l_{k})}{S_{n}-2k+5}\frac{2\Delta \deg(l_1)}{S_{n}-2k+3} \frac{2\Delta \deg(\bar{l}_{k})}{S_{n}-2k+1} \\ 
& \leq (1+o(1))\frac{2\Delta^{2}}{S_{n}} \sum_{\substack{ l_1,\dots,l_{k} \in L(\phi) \\ u,v \in \{l_1,\bar{l}_1,\dots, l_k,\bar{l}_k\}  }}
\frac{2\deg(\bar{l}_{1})\deg(l_2)}{S_{n}-1}\dots\frac{2\deg(\bar{l}_{k-1})\deg(l_{k})}{S_{n}-2k+5}\frac{2\deg(l_{1})\deg(\bar{l}_k)}{S_{n}-2k+3} \\ 
& \leq (1+o(1))\frac{2\Delta^{2}}{S_{n}} \sum_{\substack{ l_1,\dots,l_{k} \in L(\phi) \\ u,v \in \{l_1,\bar{l}_1,\dots, l_k,\bar{l}_k\}  }}\prod_{i=1}^{k}
(1+o(1))\frac{2\deg(\bar{l}_{i})\deg(l_i)}{S_{n}}\\		
& \leq (1+o(1))\frac{2\Delta^{2}}{S_{n}} \sum_{\substack{ l_1,\dots,l_{k} \in L(\phi) \\ u,v \in \{l_1,\bar{l}_1,\dots, l_k,\bar{l}_k\}  }}(1+o(1))^k\prod_{i=1}^{k}
\frac{2\deg(\bar{l}_{i})\deg(l_i)}{S_{n}}\\		
& \leq (1+o(1))\frac{2\Delta^{2}}{S_{n}} \sum_{\substack{ l_1,\dots,l_{k} \in L(\phi) \\ u,v \in \{l_1,\bar{l}_1,\dots, l_k,\bar{l}_k\}  }}(1+o(1))^k
\frac{\prod_{i=1}^{k}2\deg(\bar{l}_{i})\deg(l_i)}{S_{n}^k}\\	
& \leq (1+o(1))\frac{2\Delta^{2}}{S_{n}}(1+o(1))^k
\left(\frac{\sum_{l \in L(\phi)}2\deg(\bar{l})\deg(l)}{S_{n}}\right)^k\\	
& \leq (1+o(1))\frac{2\Delta^{2}}{S_{n}} (1+o(1))^{k}\left(\frac{2T_{n}}{S_{n}}\right)^{k}\\ 
& = (1+o(1))\frac{2\Delta^{2}}{S_{n}}\left((1+o(1))(1-\mu')\right)^{k},
\end{align*}
and the result follows.
\end{myproof}

Hence, the above two lemmas imply that $\phi$ contains no bicycles.

\begin{corollary}\label{lem:short-bicycles}
If $\frac{2T_{n}}{S_{n}}=1-\mu'<1$, then $\phi$ contains no bicycles, w.h.p.
\end{corollary}

\begin{myproof}
Let $B$ be the number of all bicycles in $\phi$, while $B_k$ is the number of bicycles of length $k$. Due to Lemma~\ref{lem:short-paths}, we expect no paths longer than $k_{0}=\lceil \frac{6}{\mu'}\log n \rceil$, which means there are no bicycles longer than $k_0$. Hence, $B = \sum_{k=2}^{k_0}B_k$. Then by Markov's inequality
\begin{align*}
\Pr\left[B > 0\right] & \leq \E B  = \sum_{k=2}^{k_0} \E B_k\\
& \leq (1+o(1))\frac{2\Delta^{2}}{S_{n}}\sum_{k=2}^{k_0} \left((1+o(1))(1-\mu')\right)^{k}, \text{ from Lemma}~\ref{lem:short-bicycles1}\\
& \leq (1+o(1))\frac{2\Delta^{2}}{S_{n}}\sum_{k=2}^{k_0} \left((1+o(1))(1-\mu')\right)^{k}
\end{align*}

Given large enough $n$, we can assume that $(1+o(1))(1-\mu')\leq 1-\frac{\mu'}{2}$. Then 		
\begin{align*}
\Pr[B > 0]  & \leq (1+o(1))\frac{2\Delta^{2}}{S_{n}}\sum_{k=2}^{k_{0}}\left((1+o(1))(1-\mu')\right)^{k}\\
			& \leq (1+o(1))\frac{2\Delta^{2}}{S_{n}}\sum_{k=2}^{k_{0}}\left(1-\frac{\mu'}{2}\right)^{k}\\
			& \leq (1+o(1))\frac{2\Delta^{2}}{S_{n}}\sum_{k=1}^{\infty}e^{-\mu'k/2}\\
			& \leq (1+o(1))\frac{2\Delta^{2}}{S_{n}}\,\frac{1}{e^{\mu'/2}-1}\\ \\
			& \leq \frac{(1+o(1))}{e^{\mu'/2}-1}\frac{2\Delta^{2}}{S_{n}}\\
			& \leq C\frac{\Delta^{2}}{S_{n}},
\end{align*}
for some constant $C>0$. Since $\alpha>2$, we recall that w.h.p. $S_n=(1+o(1))\,n\E\xi$ and $\Delta=o(n^{1/2})$. Then  
\begin{align*}
\Pr[B > 0]  & \leq C\frac{\Delta^{2}}{S_{n}} \leq C\frac{o(n)}{(1+o(1))\,n\E\xi} = o(1).
\end{align*}
Hence, w.h.p. we do not expect short bicycles in $\phi$, and, since there are no long paths, it follows that w.h.p. $\phi$ doesn't have bicycles at all.
\end{myproof}

It remains to argue that the inequality $\frac{2T_{n}}{S_{n}}=1-\mu'<1$ holds w.h.p.

\begin{proposition}
\label{lemma:alpha0_alpha}
Let $\phi \sim \mathbb{C}_{n}^{2}(\xi)$, where $\xi \sim \Class{\alpha}$ and $\E\xi^2<3\E\xi$. Then w.h.p. $\phi$ is satisfiable.
\end{proposition}

\begin{myproof}
Since $\E\xi^2$ and $\E\xi$ are both finite, we can conclude that $\alpha>2$, and so from Lemma~\ref{lemma:moments} and Theorem~\ref{theorem:sum-2}, we have that w.h.p.
\begin{align*}
S_n & =\sum_{i=1}^{n}\deg(v_i)=\sum_{i=1}^{n}\xi_i=(1+o(1)) \, n \E\xi, 
\end{align*}
and
\begin{align*}
T_n & = \sum_{i=1}^{n}d_i^{+}d_i^{-}=(1+o(1))\, n\E\left[d_i^{+}d_i^{-}\right]=(1+o(1))\, n\frac{\E\xi^2-\E\xi}{2},
\end{align*}
since $\E\left[d_i^{+}d_i^{-}\right]=\frac{\E\xi^2-\E\xi}{2}$. Hence, when $\E\xi^2<3\E\xi$ it holds
\begin{align*}
\frac{2T_n}{S_n}=(1 \pm o(1))\frac{\E\xi^2-\E\xi}{2\E\xi}=(1 \pm o(1))\left(\frac{\E\xi^2}{2\E\xi}-\frac{1}{2}\right)<1,
\end{align*}
Therefore, we can assume that
\begin{align*}
\frac{2T_n}{S_n}=1-\mu',
\end{align*}
where $0<\mu'<1$. By Lemma~\ref{lem:short-paths} there are no long paths in $\phi$, when $\E\xi^2<3\E\xi$.

Thus, it follows that there should be no \textit{long} bicycles, and by Corollary~\ref{lem:short-bicycles} there are no bicycles in $\phi$ at all.
Therefore, $\phi$ is satisfiable, which proves the proposition.
\end{myproof}

\bibliographystyle{splncs04}
\bibliography{./2sat-4.bib}
\end{document}